\documentclass[%
 reprint,
superscriptaddress,
nofootinbib,
 amsmath,amssymb,
 aps,
prd,
]{revtex4-2}

\usepackage[dvipsnames]{xcolor}
\usepackage{booktabs}
\usepackage{array}
\usepackage{tabularx}
\usepackage{adjustbox}
\usepackage{amssymb,amsmath,amsfonts}
\usepackage{dsfont} 
\usepackage{enumerate}
\usepackage{graphicx,bm}
\usepackage[hidelinks,colorlinks=true,linkcolor=blue,citecolor=blue]{hyperref}
\newcommand\aap{A\&A}                

\newcommand\apjl{ApJ}                

\newcommand\mnras{MNRAS}             

\definecolor{lzcolor}{HTML}{008B8B}

\usepackage{tabu}
\usepackage{array}
\usepackage{multirow}
\usepackage{CJK}
\usepackage{mathrsfs}
\usepackage{dcolumn}
\usepackage{mathtools}
\usepackage{graphicx}
\usepackage{dcolumn}
\usepackage{bm}
\usepackage[normalem]{ulem}

\definecolor{pumpkin}{rgb}{1.0, 0.46, 0.09}

\definecolor{pastelgreen}{rgb}{0.47, 0.87, 0.47}

\definecolor{cyan(process)}{rgb}{0.0, 0.72, 0.92}

\def\eg{\textit{e.g.}}
\begin{document}


\title{Eccentricity greatly enhances the detectability of line-of-sight acceleration in gravitational wave signals}

\author{János Takátsy}
\email{janos.takatsy@uni-potsdam.de}
\affiliation{Institut für Physik und Astronomie, Universität Potsdam, Haus 28, Karl-Liebknecht-Str. 24-25, 14476 Potsdam, Germany}

\author{Anna Puecher}
\affiliation{Institut für Physik und Astronomie, Universität Potsdam, Haus 28, Karl-Liebknecht-Str. 24-25, 14476 Potsdam, Germany}

\author{Gonzalo Morras}
\affiliation{Max Planck Institute for Gravitational Physics (Albert Einstein Institute), Am Mühlenberg 1, 14476 Potsdam, Germany}

\author{Lorenz Zwick}
\affiliation{Niels Bohr International Academy, The Niels Bohr Institute, Blegdamsvej 17, 2100 Copenhagen, Denmark}
\affiliation{Center of Gravity, Niels Bohr Institute, Blegdamsvej 17, 2100 Copenhagen, Denmark.}



\begin{abstract}
\noindent
Compact object binaries formed dynamically are characterized by a tail of eccentric sources and by the potential presence of environmental perturbations. Here we show that the presence of eccentricity in a gravitational-wave (GW) signal strongly enhances the sensitivity to environmental effects, focusing in particular on line-of-sight acceleration (LSA). We demonstrate that this enhancement is recovered in Fisher-analysis forecasts and, crucially, persists in full Bayesian parameter estimation using state-of-the-art eccentric waveform models. The key insight is that the signal should be analyzed from the moment the first detectable eccentric harmonic enters the detector band, rather than from the customary entry of the $n=2$ harmonic: this allows to probe the early inspiral and thus a longer signal, over which the GW dephasing introduced by LSA becomes substantially larger. We show that with this approach constraints on LSA can increase by more than an order of magnitude between eccentricities at 20~Hz $e_{20}=0$ and $e_{20}\sim0.1$, for signals with sufficient signal-to-noise-ratio (SNR), while we predict a factor $5$ amplification even for low-SNR ones. Fully exploiting this effect necessitates the continued improvement of efficient eccentric waveform models and likelihood evaluations, enabling the efficient analysis of signals lasting thousands of seconds and containing a large number of harmonics.
\end{abstract}

\maketitle

\section{Introduction}
\label{S:Introduction}

The first candidate eccentric gravitational-wave (GW) signals have already been identified in the LIGO--Virgo--KAGRA (LVK) catalogs \citep{Romero-Shaw:2020thy,Romero-Shaw:2021ual,Romero-Shaw:2022xko,Iglesias:2022xfc,Romero-Shaw:2025vbc,Gupte:2024jfe,Morras:2025xfu,Jan:2025fps,Kacanja:2025kpr,McMillin:2025hof,Planas:2025plq,Tiwari:2025fua}. The current picture that binary systems of stellar-mass compact objects originate from a mixture of isolated and dynamical formation channels \citep{Mandel:2018hfr,Mapelli:2020vfa,Mandel:2021smh,Zevin:2021rtf,2024stegmann,StegmannKlencki2025} suggests that a fraction of detected signals must retain some amount of eccentricity in the detector band \citep{2006ApJ...640..156G, 2017ApJ...840L..14S,silsbee2017,antonini2017, Samsing18a, 2018ApJ...855..124s,
2018MNRAS.tmp.2223s, 2018PhRvD..98l3005R, 2019ApJ...881...41L, 2019PhRvD.100d3010S, 2020PhRvD.101l3010S,2019PhRvD..99f3003K,Romero-Shaw:2021ual,OLeary:2008myb,Antonini:2012ad,Samsing14,Gondan:2017wzd,Samsing:2017xmd,Takatsy:2018euo,Zevin:2018kzq,Rasskazov:2019gjw,Tagawa:2019osr,Fragione:2019vgr,Tagawa:2020jnc,Samsing:2020tda,DallAmico:2023neb}. It is therefore likely that more will emerge in future observing runs, in particular as detector sensitivities improve toward lower frequencies. Beyond their astrophysical significance, eccentric GW signals present some additional opportunities for parameter inference. The presence of eccentric harmonics extends the observable inspiral to earlier times and also provides multiple, independently measurable phase evolutions, improving constraints on both intrinsic and extrinsic binary parameters \citep{Yunes:2009yz,Mikoczi:2012qy,Sun:2015bva,Moore:2019vjj,Sinha:2025vmc,Pompili:2026yxq}. Recently, it was shown that these same properties can substantially enhance the measurement of environmental signatures encoded in the GW phase \citep{Takatsy:2025bfk,Zwick:2025qzv}. Collectively known as ``environmental effects'', GW perturbations caused by acceleration, drag and tidal forces can provide direct information about the surroundings of the binary \citep{1993chakrabarti,1995ryan,2008barausse,2007levin,Kocsis:2011ka,2014barausse,2017inayoshi,2017meiron,2017Bonetti,2019alejandro,2019randall,2020cardoso,DOrazioGWLens:2020,2022liu,2022xuan,garg2022,2022cole,2022chandramouli,2022sberna,2023zwick,2023Tiede,2021alejandro,2022destounis,2022cardoso,2020caputo,2022zwick,Vijaykumar2023-qg,2024zwicknovel,2025Zwick_ecc,2021andrea,2024basu,2024santoro,2025vicente,2025dyson,2025destounis,2025copparoni,2025torres,Zwick:2025wkt,Tagawa:2025tfd}. Line-of-sight acceleration (LSA) offers one of the cleanest examples \citep{Yunes:2010sm,Bonvin:2016qxr,2017meiron,2019robson,Tamanini2020,Vijaykumar2023-qg,Samsing:2024syt,Hendriks:2024gpp,Hendriks:2026kys,Tiwari:2023cpa,Tiwari:2024pvb,Lazarow:2024gdn,Tiwari:2025aec,Gera:2025ugl,Roy:2026duh,LIGOScientific:2026fcf}, producing a characteristic negative post-Newtonian correction to the GW phase. Such accelerations can arise naturally in formation channels involving close encounters and prompt GW captures, as well as in active galactic nuclei disks \citep{Samsing:2025phj,Tagawa:2025tfd}. In the case of eccentric binaries, preliminary work has suggested that the resulting dephasing is significantly amplified by the higher-harmonic structure of the signal \citep{Takatsy:2025bfk,Zwick:2025qzv}, suggesting a substantial increase in detectability compared to circular binaries.

In this paper, we delineate the properties of eccentric signals that lead to an enhancement of the detectability of line-of-sight acceleration effects. We then show that this enhancement persists in both Fisher-matrix forecasts and, crucially, in full Bayesian parameter estimation (PE) using state-of-the-art eccentric waveform models and sources representative of the LVK population. 
We propose the following data-analysis strategy aimed at exploiting this enhancement in eccentric signals more effectively: The key requirement is to analyze the signal from the moment the first detectable harmonic enters the detector band, rather than from the customary entry of the $n=2$ harmonic.
In the time domain, this has the following simple interpretation. For an eccentric binary the GW frequency is not monotonic, but peaks at each periastron passage, where the instantaneous frequency exceeds the orbit-averaged one, typically used by waveform models to define the starting frequency. Therefore, the periastron frequency can reach the lower edge of the detector band much earlier than the orbit-averaged frequency for eccentric systems, allowing us to probe the early inspiral of the binary. This point, and its impact on PE, was also made in Refs.~\cite{Sinha:2025vmc,Pompili:2026yxq}.
Here we also show how not adopting the above requirement explains why recent studies jointly inferring eccentricity and line-of-sight acceleration have not recovered the expected enhancement \cite{Roy:2026mco,Pompili:2026tdf,Ansari:2026nhb}.

The paper is organized as follows: in Sec.~\ref{S:Methods} we explain the enhancement of LSA dephasing in eccentric signals, and delineate the analysis methods employed for both the full Bayesian inference and the Fisher matrix analysis. We present the results for the analysis of a set of injections, and for the real event GW200105\_162426 (GW200105) in Sec.~\ref{S:Results}, and we summarize our conclusions in Sec.~\ref{S:Discussion}. We also provide additional derivations and results for further tests in the appendix.

\section{Methods}
\label{S:Methods}

As shown in Refs.~\cite{Zwick:2025wkt,Takatsy:2025bfk,Zwick:2025qzv}, the enhancement of dephasing signatures in eccentric
binaries originates from a simple argument. For circular orbits, an environmental dephasing typically scales as a negative power of the GW frequency $f$,
\begin{equation}
    \Delta\Psi_{\rm circ}^{n=2} \propto f^{\alpha} \:,
    \qquad \alpha<0 \: .
\end{equation}
In an eccentric binary, the $n$-th eccentric GW harmonic observed at a frequency $f$ is emitted when the binary orbital frequency is
$F=f/n$. Consequently, the corresponding phase shift scales as:
\begin{equation}
    \Delta\Psi_n(f)
    \propto n\,
    \Delta\Psi_{\rm circ}^{n=2}(f)\,n^{-\alpha} \left[ \frac{t_{\rm insp}(e,F)}{t_{\rm insp}(e=0,F)}\right]^2 \: ,
\end{equation}
where $t_{\rm insp}$ is the inspiral time at orbital frequency $F$, and the additional factor of $n$ comes from the increased phase accumulation in higher harmonics.\footnote{Note the difference between the eccentric harmonic number $n$, corresponding to the Fourier decomposition of the GW in terms of the mean anomaly, and the spherical harmonics $(l,m)$ that correspond to the spatial spherical decomposition used for the GW higher-order modes. Also note that while later in the text we often refer to specific modes by their eccentric harmonic index $n$ alone, all three indices need to be specified in order to uniquely identify modes. The other two indices are also explicitly stated whenever needed for clarity.}
Physically, this result reflects the fact that higher harmonics probe progressively earlier stages of the inspiral, where environmental effects are intrinsically stronger. For LSA, $\alpha=-13/3$, therefore the contribution from the $n$-th harmonic to the environmental signal-to-noise ratio, defined using the difference strain due to LSA, $\mathrm{SNR}^\mathrm{LSA}\propto \Delta h \equiv h_\mathrm{LSA} - h_\mathrm{noLSA}$, is:
\begin{equation}
    {\rm SNR}^{\rm LSA}_n\sim n^{16/3} \, \mathrm{SNR}_n \: ,
\end{equation}
leading to an expected enhancement in ${\rm SNR}_{\rm LSA}$ of:
\begin{equation}
    {\rm SNR}^{\rm LSA}\sim n_{\rm max}^{16/3} \, \mathrm{SNR}_{n_\mathrm{max}} \:,
\end{equation}
where $n_{\rm max}$ denotes the highest harmonic that contributes appreciable power to the observed signal. Crucially, the large power power of $16/3$ means that even including few additional eccentric harmonics will significantly enhance the detectability of the dephasing induced by LSA.

The purpose of this section is to show how this effect can be rigorously implemented in GW waveforms and to provide details of our Bayesian inference using a state-of-the-art eccentric waveform model, \texttt{pyEFPEHM} \cite{Morras:2025nlp,Morras:2026fho}, as well as for the broader parameter scan with Fisher analysis.

\subsection{Phase shift from line-of-sight acceleration}

We parametrize the LSA effects by $\Gamma\equiv a_\parallel/c$, i.e.,  the acceleration of the center-of-mass of the binary along the line of sight divided by the speed of light (in units of ${\rm s^{-1}}$), taken to be positive when the acceleration points away from the observer.
To first order in $\Gamma$, the effect of the LSA can be included in frequency domain waveforms through the phase shift \cite[\eg][]{Bonvin:2016qxr,Vijaykumar2023-qg,Samsing:2024syt,Takatsy:2025bfk,Roy:2026mco}:
\begin{equation}
    \Delta\psi_j(f) = \pi \Gamma f t_j(f)^2 \: ,
    \label{eq:DPsi_LSA}
\end{equation}
where $t_j(f)$ is the stationary time, measured from coalescence, at which the harmonic $j$ has frequency $f$. The label $j=(l,m,n)$ denotes the spherical-harmonic multipole $(l,m)$ and the eccentric harmonic $n$ of the mean anomaly.

One important observation to make here is that, for a fixed GW frequency, the dependence of Eq.~\eqref{eq:DPsi_LSA} on the harmonic is entirely encoded in the stationary time. Harmonics with large $n$ reach a given frequency at increasingly earlier times in the inspiral. Since the LSA-induced dephasing grows as $t^2$, it is much larger for modes with large $n$. This is the origin of the enhanced sensitivity of eccentric harmonics to environmental effects. Additional corrections to the GW amplitude are suppressed by a factor of $(t_jf)^{-1}$ (see Appendix~\ref{app:high_order_LSA}), hence they are not included in this study.

\subsection{Bayesian inference using \texttt{pyEFPEHM}}

In the full Bayesian PE analysis, the parameters $\vec{\theta}$ of a system emitting GWs are inferred through Bayes theorem:
\begin{equation}
    p(\vec{\theta}|d, \mathcal{H}) = \frac{p(d|\vec{\theta}, \mathcal{H}) p (\vec{\theta}|\mathcal{H})}{p(d | \mathcal{H})}.
    \label{eq:bayes}
\end{equation}
In Eq.~\eqref{eq:bayes}, $p(\vec{\theta}|d, \mathcal{H})$ denotes the \textit{posterior} probability of a set of parameters $\vec{\theta}$ given the data $d$ and a model (or hypothesis) to describe the data $\mathcal{H}$, $p(d|\vec{\theta}, \mathcal{H})$ is the \textit{likelihood} of the data, and $p(\vec{\theta}|\mathcal{H})$ the \textit{prior} probability of the parameters $\vec{\theta}$ given $\mathcal{H}$, which encodes our a-priori knowledge about the system. 
In PE analyses, the evidence $p(d|\mathcal{H})$ at the denominator serves just as a normalization constant.

To describe the data we employ the waveform model \texttt{pyEFPEHM}~\cite{Morras:2026fho}, an inspiral-only approximant including eccentricity, precession, and higher-order modes. It extends \texttt{pyEFPE}~\cite{Morras:2025nlp} within the efficient fully precessing eccentric (EFPE) framework~\cite{Klein:2018ybm,Klein:2021jtd,Arredondo:2024nsl}. The orbital motion is described with the quasi-Keplerian parametrization and the spin precession with a multiple-scale analysis, so that only the precession-averaged radiation-reaction equations need to be integrated numerically. These combine eccentric post-Newtonian (PN) results (3PN for aligned spins) with quasi-circular corrections up to 4.5PN order and horizon absorption~\cite{Alvaro-Diaz:2026acu}. All $(l,m)$ multipoles entering the GW amplitude up to 1PN order are included, and each $(l,m)$ multiple is analytically decomposed into harmonics $n$ of the mean anomaly, with amplitudes valid at arbitrary eccentricity~\cite{Morras:2025nbp}. The frequency-domain waveform is computed harmonic by harmonic with the stationary phase approximation, using shifted uniform asymptotics for the precession-induced modulations~\cite{Klein:2014bua}. The model thus provides the stationary time $t_j(f)$ of every harmonic $j=(l,m,n)$, measured from coalescence, which allows us to apply Eq.~\eqref{eq:DPsi_LSA} mode by mode at a negligible additional cost.

For a few selected values of eccentricity, we perform PE with \textsc{Bilby\_pipe}~\cite{Ashton:2018jfp, Romero-Shaw:2020owr} and the nested sampler \texttt{dynesty}~\cite{dynesty}, using 2048 live points. To make the analysis computationally feasible, we speed-up the likelihood evaluation by using the multibanding approximation~\cite{Morisaki:2021ngj}, which exploits the chirping nature of GW signals: the data strain is divided into frequency bands, and the likelihood is evaluated on custom frequency grids, determined by the specific time to merger of each band. We use multibanding with highest-mode 6 and an accuracy factor of 10. More details about the reliability of multibanding in the analysis of eccentric signals are provided in Appendix~\ref{sec:mb_tests}.

\subsection{Analysis setup}
\label{ssec:setup}

We simulate signals in a detector network comprising two Advanced LIGO in Hanford and Livingston~\cite{LIGOScientific:2014pky}, and Advanced Virgo~\cite{VIRGO:2014yos}. To focus on the effects under investigation and avoid potential biases due to noise fluctuations, we run in a zero-noise configuration.
We analyze the signals in the frequency range [20, 100]~Hz, where the upper limit is determined by the fact that the waveform needs to be terminated before the innermost stable circular orbit (ISCO). As discussed above, the effects we are interested in are expected at low frequencies, therefore the upper cutoff does not risk discarding any information. 
Crucially, even if the analysis is started at 20~Hz (given the detector sensitivity limits), the waveform generation, both for the injection and during the sampling process, is initiated at the point where the frequency of the 22-mode $n=2$ harmonic is $f_{\rm ini}$. This ensures that harmonics up to $n_\mathrm{max}=(40~\mathrm{Hz})/f_{\rm ini}$ start below 20~Hz, meaning that all the relevant information from these harmonics is included in the model.

For both Fisher and PE analyses, we consider a non-spinning source with equal masses $m_1 = m_2 = 20M_\odot$.
This choice of relatively large masses helps reduce the computational cost of the PE analysis, while still ensuring that the signal has enough cycles in band. 
For the PE injections, we place the source at a fixed random sky location, and at a luminosity distance of $100$~Mpc, resulting in an SNR $~\sim 110-120$ depending on the system's properties. For all the tests described below, we set $\Gamma = 0$.

In Appendix~\ref{sec:further_tests}, we investigate the robustness of the analysis in case of different systems, such as precessing sources, as well as unequal-masses, and $\Gamma$ different from zero.

We employ priors uniform in component masses, with chirp mass sampled in the range $[17.0, 18.0] \, M_\odot$ and mass ratio in $[0.05, 1.0]$. Eccentricity and mean anomaly are sampled uniformly in $[0.0, 0.4]$ and $[0, 2\pi]$, respectively, while for $\Gamma$ we choose a uniform prior in $[-0.01, 0.01]$. 
The luminosity distance prior is uniform in source frame, in the range $[10,1000]$~Mpc, and for all the other parameters we use the standard uninformative priors implemented in \textsc{Bilby} and described in Ref.~\cite{Romero-Shaw:2020owr}.

\subsection{Fisher analysis}

In order to be able to scan a large portion of the parameter space and understand the general trends in the measurement uncertainty of the LSA in an efficient way, we also study eccentric systems using Fisher analysis \cite{Finn:1992wt,1994cutlerflanagan}. For this approach we also use \texttt{pyEFPEHM} together with the same setup as described in Sec.~\ref{ssec:setup} and $\Delta\psi_n(f)$ calculated according to Eq.~\eqref{eq:DPsi_LSA}.

The Fisher matrix is defined as:
\begin{equation}
    \mathcal{F}_{\vartheta\vartheta'} = \left(\partial_\vartheta h | \partial_{\vartheta'} h\right) \: ,
\end{equation}
where $\vartheta$ denotes the waveform parameters and $(\cdot|\cdot)$ is the usual noise-weighted inner product. The derivatives with respect to the various parameters are calculated numerically, without any analytic shortcuts that might not hold with higher modes, eccentricity or precession. Each one is computed in three independent runs of central finite differences (one-sided at the boundary of the domain of validity), using symmetric steps and Richardson extrapolation up to fourth order. The initial step is chosen by a pilot search such that the change in the waveform is linear. The three runs are then combined through taking the median in each frequency bin. To ensure the waveform is smooth in the parameters, the adaptive harmonic selection of \texttt{pyEFPEHM} is frozen at the injection. This is especially important when calculating the derivatives at low eccentricities, where the adaptive harmonic selection can make these derivatives non-convergent.

We then calculate the measurement uncertainty of $\Gamma$ marginalized over the remaining parameters using the inverted Fisher matrix:
\begin{equation}
    \sigma_\Gamma = \sqrt{(\mathcal{F}^{-1})_{\Gamma\Gamma}} \: .
    \label{eq:Fisher_marg_sigma}
\end{equation}

\section{Results}
\label{S:Results}

We investigate how the constraints on $\Gamma$ change with the system eccentricity and with the starting frequency at which the waveform for the analysis is generated. We look at a few selected configurations with full PE, and then investigate the trend systematically with Fisher matrices.
\begin{figure*}[tbh]
        \centering        
        \includegraphics[width=0.85\textwidth]{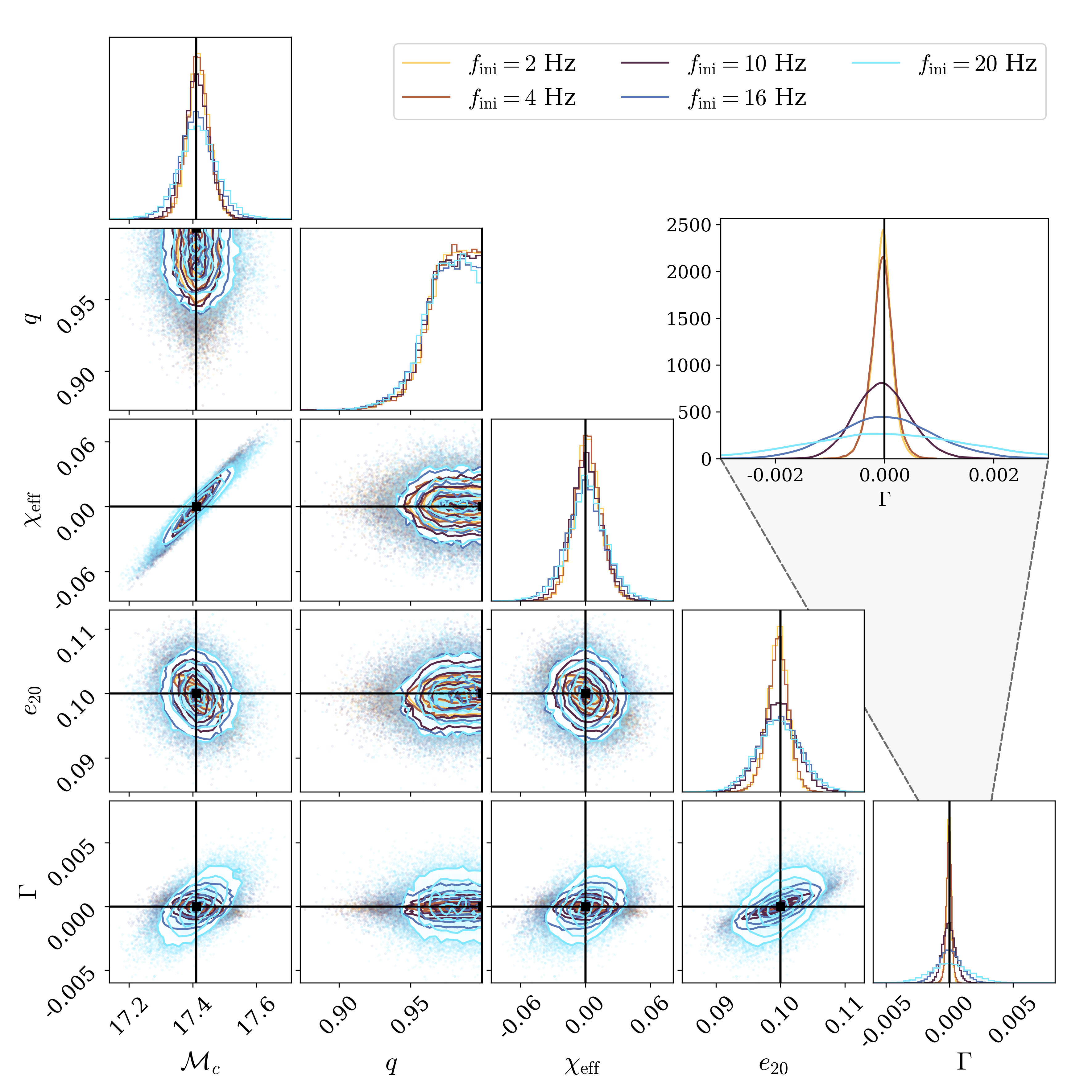}
    \caption{Posterior probability densities for chirp mass $\mathcal{M}_c$, mass ratio $q$, effective inspiral spin $\chi_{\rm eff}$, eccentricity value at 20~Hz $e_{20}$, and line-of-sight acceleration parameter $\Gamma$ for analyses where the model is generated starting form different initial frequencies $f_{\rm ini}$. The inset zooms on the posterior distribution of $\Gamma$. The black vertical lines indicate the injected values.}
    \label{fig:fini_diff}
\end{figure*}

\begin{figure*}[tbh]
        \centering        
        \includegraphics[width=0.85\textwidth]{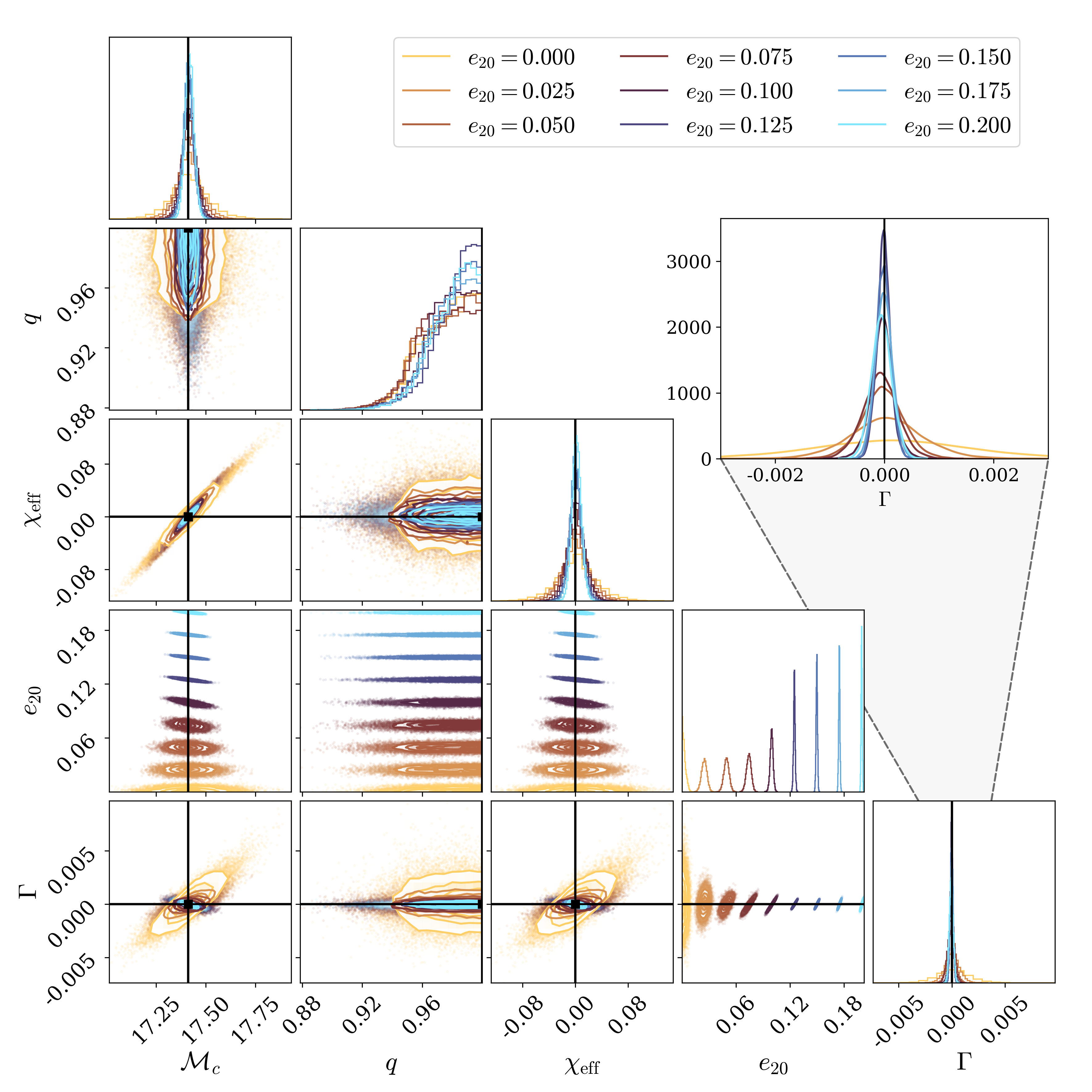}
    \caption{Posterior probability densities for chirp mass $\mathcal{M}_c$, mass ratio $q$, effective inspiral spin $\chi_{\rm eff}$, eccentricity value at 20~Hz $e_{20}$, and line-of-sight acceleration parameter $\Gamma$ for signals simulated from systems with different eccentricity values. The inset zooms on the posterior distribution of $\Gamma$. The black vertical lines indicate the injected values.}
    \label{fig:ecc_diff}
\end{figure*}

\subsection{Results from Bayesian inference}

\subsubsection{Varying initial waveform frequency}

As discussed in Sec.~\ref{S:Methods}, the phase-shift contribution of LSA scales with a large negative power of frequency, hence most information about this effect is imprinted in the early evolution of the system. In eccentric systems, we can access this early stage thanks to higher eccentric harmonics that enter the band earlier than the dominant quasi-circular mode. In Bayesian inference, at each sampling step the data is compared to the GW signal that would be generated by the sampled parameters with the chosen waveform model. The initial frequency $f_{\rm ini}$ at which we start generating the waveform plays a crucial role: the lower the frequency, the more information from higher harmonics, produced at earlier times, appears in band. Therefore, a sufficiently low $f_{\rm ini}$ is required to fully exploit the enhancement. As an example, if $f_{\rm ini}=20$~Hz, then only the full information from the $n=1,2$ harmonics is recovered in the detector frequency band. Only a fraction of the $n\geq3$ would be included in band, since (\textit{i}) their contribution to the signal would start at a GW frequency $f_{\rm GW}=30,40,...$~Hz, while (\textit{ii}) the early evolution of the binary where these harmonics would have a much stronger contribution is completely ignored. Instead, with $f_{\rm ini}=2$~Hz, all the measurable information from up to the $n_{\rm max}=20$ harmonics is fully included in the waveform.

Fig.~\ref{fig:fini_diff} shows the posterior probability distributions resulting from our Bayesian inference on signals simulated as described in Sec.~\ref{ssec:setup}, with eccentricity at 20~Hz set to $e_{20} = 0.1$. The figure compares results from analyses with different waveform $f_{\rm ini}$, showing how the constraints on eccentricity and especially on $\Gamma$ become tighter with lower $f_{\rm ini}$. Notably, the constraint remains unchanged when $f_{\rm ini}$ goes from 4~Hz to 2~Hz: for the chosen eccentricity value $e_{20}=0.1$, a waveform generated from 4~Hz already contains all the relevant eccentric harmonics information. All further harmonics would have $\mathrm{SNR}^n\ll1$, therefore lowering $f_{\rm ini}$ yields no significant improvement (see discussion in Sec.~\ref{SS:Fisher_results}). Larger values of eccentricity will require lower values of $f_{\rm ini}$ to encode all the harmonic content, and consequently the information we can obtain from them also saturates at lower $f_{\rm ini}$ values, as shown later in Fig.~\ref{fig:fisher_sigma}.

\subsubsection{Varying eccentricity}
\label{SSS:Bayes_ecc}

Fig.~\ref{fig:ecc_diff} shows the posteriors obtained for analyses of systems with different eccentricities, while everything else is kept fixed as described in Sec.~\ref{ssec:setup}. Here we use $f_{\rm ini} = 4 \, {\rm Hz}$ for all the analyses. As expected, more eccentric systems lead to overall tighter constraints also on $\Gamma$. However, this trend inverts after $e=0.125$. For increasing eccentricity, more and more higher harmonics start contributing to the signal, which then carry information about the early evolution of the binary, providing more stringent constraints on $\Gamma$. However, if the value of $f_{\rm ini}$ is fixed, the bound on the length of the signal cuts off information carried by higher harmonics from the very early inspiral, and hence constraints on $\Gamma$ saturate. At very high eccentricities, the inspiral time and thus the signal duration start decreasing, naturally giving worse constraints. Overall, the two competing effects from a richer harmonic content and a shorter overall duration generate an optimal eccentricity for which $\Gamma$ can be measured with the smallest uncertainty.

Note that the measurement accuracy of $\Gamma$ is improved by slightly more than an order of magnitude between systems with $e_{20}=0$ and $e_{20}=0.125$, which can have profound implications on the observability of LSA and of environmental effects in general in mildly-eccentric systems.

\subsection{Results from Fisher analysis}
\label{SS:Fisher_results}

\subsubsection{Fisher information by harmonic}

\begin{figure}[t]
    \centering        
    \includegraphics[width=\columnwidth]{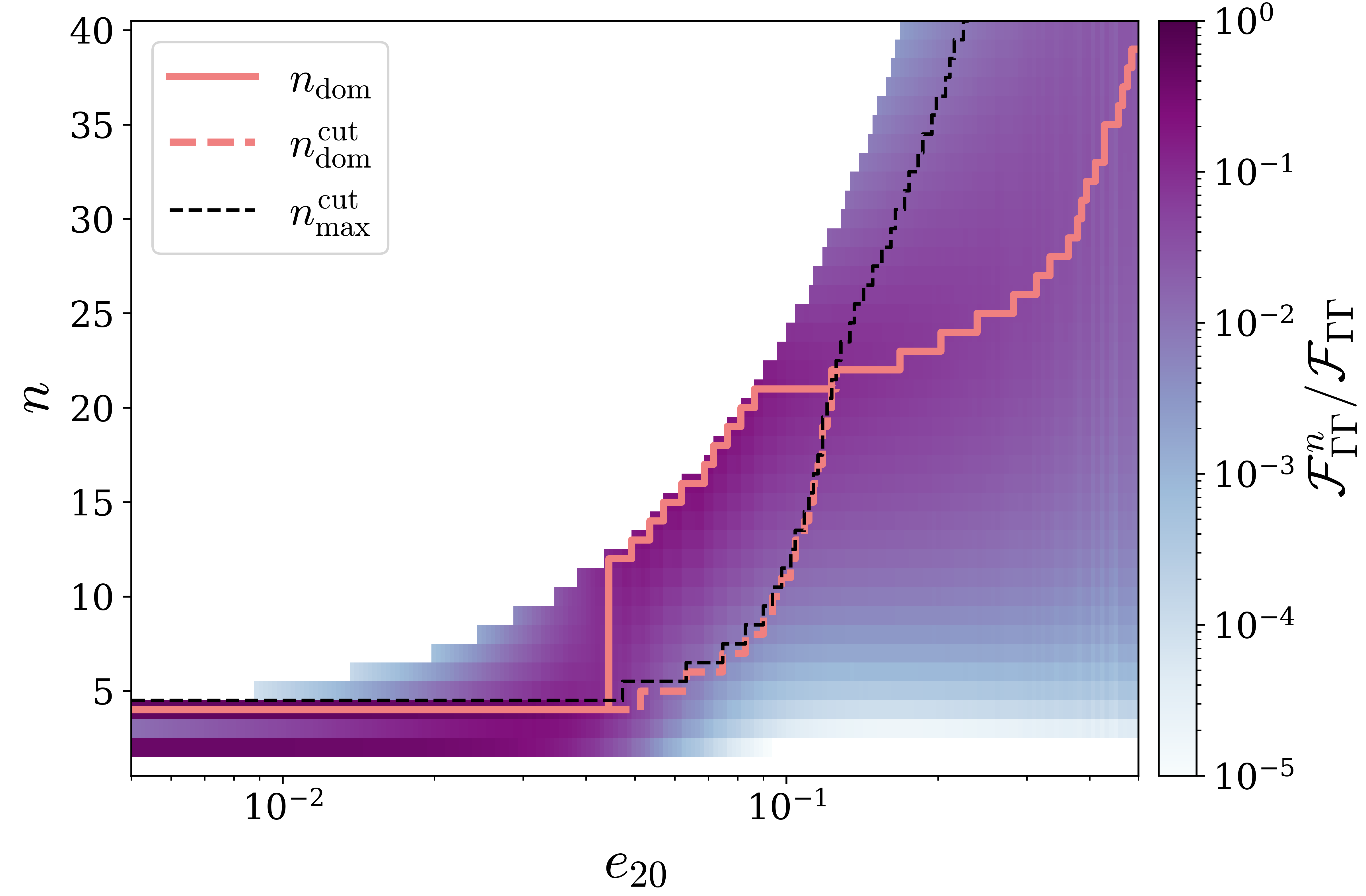}
    \caption{Relative contribution to the Fisher matrix element $\mathcal{F}_{\Gamma\Gamma}$ from different harmonics, as a function of orbital eccentricity $e_{20}$. The solid line denotes the harmonic with the dominant contribution at each eccentricity value. The black dashed line show the boundary above which harmonics are neglected after applying the SNR cut for harmonics with $(\mathrm{SNR}^n)^2<0.5$, while the red dashed line corresponds to the dominant harmonics after this cut. The waveforms were generated with $f_{\rm ini}=2$~Hz.}
    \label{fig:fisher_ndom}
\end{figure}

In order to first gain insight about the contribution of various harmonics to the enhancement in measurement precision of $\Gamma$, it is helpful to look at the contribution to the Fisher-matrix element $\mathcal{F}_{\Gamma\Gamma}$ from the different harmonics, at various eccentricities. This matrix element still misses information about parameter correlations, but can point us to the main factors contributing to the enhancement. Fig.~\ref{fig:fisher_ndom} shows, for $f_{\rm ini}=2$~Hz, the ratio $\mathcal{F}^n_{\Gamma\Gamma}/\mathcal{F}_{\Gamma\Gamma}$, where $\mathcal{F}_{\Gamma\Gamma}$ is computed from the full waveform and 
\begin{equation}
    \mathcal{F}^n_{\Gamma\Gamma}=\sum_{l,m}(\partial_\Gamma h_{lmn}|\partial_\Gamma h_{lmn}) \: ,
\end{equation}
summing over all the $(l,m)$ multipoles with the same $n$. The cross terms between harmonics are thus neglected in $\mathcal{F}^n_{\Gamma\Gamma}$, and $\sum_n\mathcal{F}^n_{\Gamma\Gamma}$ only approximates $\mathcal{F}_{\Gamma\Gamma}$. Nevertheless, these cross terms are small, since harmonics with different $n$ reach a given frequency at different times in the inspiral, and for harmonics with the same $n$ but different $m$, periastron advance makes their relative phase wind many times across the band, so in both cases they average out. Only multipoles with the same $(m,n)$ add coherently, but the higher-$l$ ones are much weaker.

The solid red line shows the harmonic with the dominant contribution to the Fisher-matrix element. For $e_{20}\lesssim 10^{-2}$, even though contributing to roughly half of the Fisher-matrix element, most of the information comes not from the $n=2$ harmonic, as one would naively expect, but from an $n=4$ harmonic. This is because even though in our equal-mass system higher-order modes are suppressed, the fact that the dominant harmonic from the $(l,m)=(4,4)$ mode carries information from an earlier epoch, greatly amplifies its importance. The $n=4$ harmonic of the $(2,2)$ mode, instead, scales as $e^2$ and is negligible at these eccentricities. For eccentricities as low as $e_{20}\sim 10^{-2}$ a few higher harmonics start giving relevant contribution. At eccentricity $e_{20}\sim4\times 10^{-2}$, when the eccentricity is high enough such that in the early phases of the evolution a significant number of higher harmonics would appear with a substantial contribution, the dominant harmonic suddenly jumps up to the maximum harmonic included in the waveform, then climbs up to reach the $n=20$ harmonic, only going further up for $e_{20}\gtrsim0.1$. This saturation behaviour at $n=20$ is exactly due to the somewhat artificial choice of initializing the waveform at $f_{\rm ini}=2$~Hz, which means that the highest harmonic that has its full observable contribution in the waveform is $n_\mathrm{max}=2\times(20~\mathrm{Hz}/2~\mathrm{Hz})=20$. Higher harmonics might potentially have more information content but this information is stored in a part of the waveform corresponding to $f_{\rm ini}<2$~Hz.

The sharp jump of the dominant harmonic at $e_{20}\sim4\times 10^{-2}$, which would reach even higher harmonics provided that they were included in the waveform is due to a limitation of the Fisher-matrix approach. Since the Fisher matrix approximates waveform differences from local derivatives, it does not take into account the saturation effect, i.e., the fact that the difference in strain for phase shifts $\Delta\psi\gtrsim1$ do not grow linearly but are bounded form above. Due to this, the contribution of harmonics with a low SNR but substantial phase shift at early times are amplified beyond their real effect on the measurement uncertainty.

We solve this problem heuristically by dropping harmonics with an SNR below a threshold $(\mathrm{SNR}^n_{\rm cut})^2=0.5$. This choice of value reflects the fact that the contribution of a harmonic with $(\mathrm{SNR}^n_{\rm cut})^2\leq0.5$ to the log-likelihood is $\leq 1$, even with a maximal dephasing. Thus, these harmonics, even with a maximal dephasing, cannot exclude a $1\sigma$ change of $\Gamma$. Note that such a sharp cut in harmonics with low SNR can create discontinuous features in the measurement uncertainties of $\Gamma$, since the contributions of harmonics with SNR growing above this threshold would change from zero to a finite, non-negligible value. In a full PE setup, however, the contribution from harmonics below this threshold would not be neglected completely but would be bounded, leading to a smooth dependency on the eccentricity. In addition, we emphasize that the exact choice of the SNR threshold remains heuristic. Therefore, developing more robust approaches to handling low-SNR harmonics in Fisher analyses remains an important direction for the future.

The black dashed line in Fig.~\ref{fig:fisher_ndom} shows the boundary above which harmonics are dropped due to the SNR cut discussed above. This suppresses unphysical contributions from higher harmonics between $e_{20}\sim0.04-0.1$, leading to a smooth increase of the dominant harmonic in this range (shown by the red dashed line). The saturation of the dominant harmonic above this range indicates that even tighter LSA constraints could be gained at high eccentricities by choosing an even lower value of $f_{\rm ini}$.

\subsubsection{Forecast on marginalized uncertainties}

\begin{figure}[t]
    \centering        
    \includegraphics[width=\columnwidth]{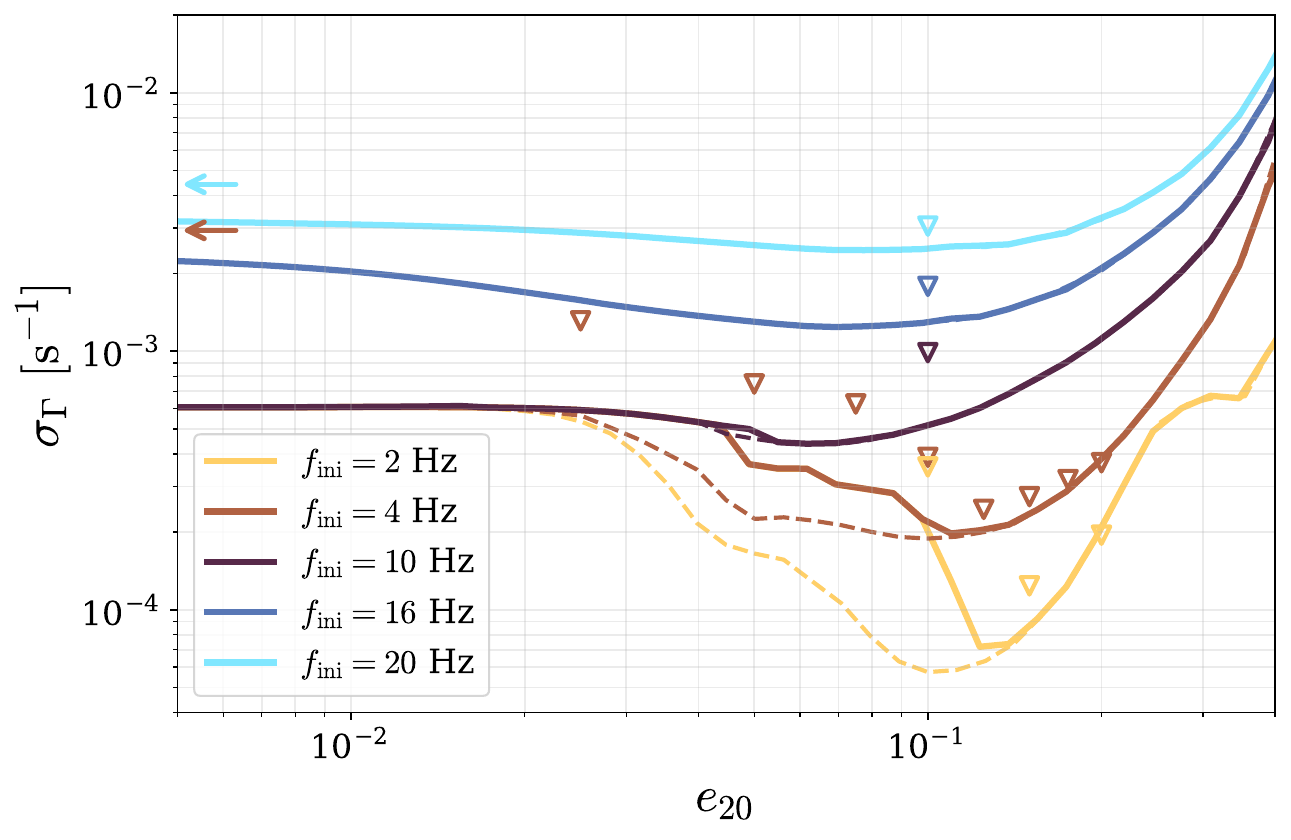}
    \caption{The measurement error of $\Gamma$ as a function of the eccentricity for different values of $f_{\rm ini}$. Results from the Fisher analysis without the SNR cut are depicted by the dashed lines, while the ones including the SNR cut are shown by the solid lines. Results from full PE for selected parameters are shown by markers. The zero-eccentricity PE results for $f_{\rm ini}=20$~Hz and 4~Hz are denoted by arrows.}
    \label{fig:fisher_sigma}
\end{figure}

The measurement error of $\Gamma$ marginalized over the other parameters, $\sigma_\Gamma$, is calculated according to Eq.~\eqref{eq:Fisher_marg_sigma}.
Fig.~\ref{fig:fisher_sigma} shows this error as a function of eccentricity $e_{20}$ for different values of $f_\mathrm{ini}$ shown by lines with different colors, while markers represent the errors obtained from full PE. Dashed lines correspond to Fisher results without the SNR cut, while the solid lines represent results with this cut included.

The two types of trends discussed before in varying eccentricity and $f_{\rm ini}$ are both visible in Fig.~\ref{fig:fisher_sigma}, i.e. that (\textit{i}) choosing a lower value for $f_{\rm ini}$ provides more in-band information from the early inspiral carried by higher harmonics, thus giving tighter constraints, which for low eccentricities already saturates for $f_{\rm ini}\sim4-10$~Hz; and that (\textit{ii}) choosing a sufficiently low value for $f_{\rm ini}$ and increasing the eccentricity enhances constraints on $\Gamma$ substantially, which then start degrading at high values of eccentricity. Note that there is some improvement in $\sigma_\Gamma$ even at zero eccentricity due to the higher-order spherical modes accounted for in \texttt{pyEFPEHM}.

The Fisher analysis, being an approximation, naturally underestimates the measurement uncertainties, however, the general trends are well-recovered. The main differences between the PE and the Fisher results are the difference in improvement at $e_{20}\sim0$, and at eccentricities in the range $e_{20}\sim0.04-0.1$ between full PE and Fisher results without the SNR cut. The second is due to the effect of the Fisher analysis overestimating the contribution from low-SNR harmonics, and is in fact corrected once the SNR cut is introduced. The first is also due to a similar reason: the SNR of the $(l,m)=(4,4)$ mode contribution does not fall below the threshold and thus its contribution, which would be reduced in a full PE setup, is fully accounted for in the Fisher analysis. Note that this large difference between PE and Fisher is especially true for lower eccentricities, where including or not including a harmonic in addition to the other $1-2$ harmonics can make a huge difference in the results, while this effect at higher eccentricities with $\sim10$ harmonics already included is more subtle. Notably, $\sigma_\Gamma$ matches the PE results quite well at $e_{20}\gtrsim0.1$ and $f_{\rm ini}\leq4$~Hz, and for these $f_{\rm ini}$ values we also recover the location of the optimal eccentricity relatively accurately. As discussed previously, while results from the Fisher analysis without the SNR cut show a smooth dependency on the eccentricity, the introducing the sharp SNR cut creates discontinuous breaks in this trend.

The increase of $\sigma_\Gamma$ at high eccentricities is recovered in the Fisher results as well and is due to the reasons discussed in Sec.~\ref{SSS:Bayes_ecc}, i.e. that such high eccentricities substantially reduce the inspiral time of the binary, thus reducing the phase shift as well.

As mentioned before, the errors for $\Gamma$ shown here were marginalized in order to take into account correlations with other parameters, as they can increase the value of $\sigma_\Gamma$ by more than an order of magnitude; in particular, we find that the effect of these correlations is the strongest for $e_{20}\gtrsim0.1$.

\subsubsection{Comparison with recent results}

Multiple recent studies investigated constraints on LSA using eccentric waveforms \cite{Roy:2026mco,Pompili:2026tdf,Ansari:2026nhb,Ansari:2026pxp,Pathak:2026cik}. Notably, Refs.~\cite{Roy:2026mco,Pompili:2026tdf,Ansari:2026nhb} found that the value of $\Gamma$ is recovered with a larger uncertainty for binaries with eccentricity $e_{20}\sim0.2-0.4$ compared to the circular case. In contrast, we find that introducing eccentricity to the binary system can greatly enhance the measurability of $\Gamma$. For $f_{\rm ini}=20$~Hz, however, even though $\sigma_\Gamma$ slightly decreases between $e_{20}=0$ and $e_{20}=0.1$, it increases significantly above $e_{20}\sim0.1$. While for eccentricities $e_{20}\gtrsim0.5$ we also predict the accuracy of LSA-recovery to degrade for all values of $f_{\rm ini}$, the worse recovery of $\Gamma$ for eccentric binaries reported in Refs.~\cite{Roy:2026mco,Pompili:2026tdf} is almost certainly linked to the reason discussed before, i.e. to initializing the waveform at a higher reference frequency and thus ignoring the early evolution of the binary, where most of the information from the LSA carried by higher harmonics would come from. In fact, it is directly implied in Ref.~\cite{Pompili:2026tdf} that all waveforms were initialized at $f_{\rm ini}=20$~Hz, while a fixed signal duration of 256~s is mentioned in Ref.~\cite{Ansari:2026nhb}, once again implying signal generation from $20$~Hz for a GW170817-like system.

\subsubsection{Dependence on signal-to-noise ratio}

\begin{figure}[t]
    \centering        
    \includegraphics[width=\columnwidth]{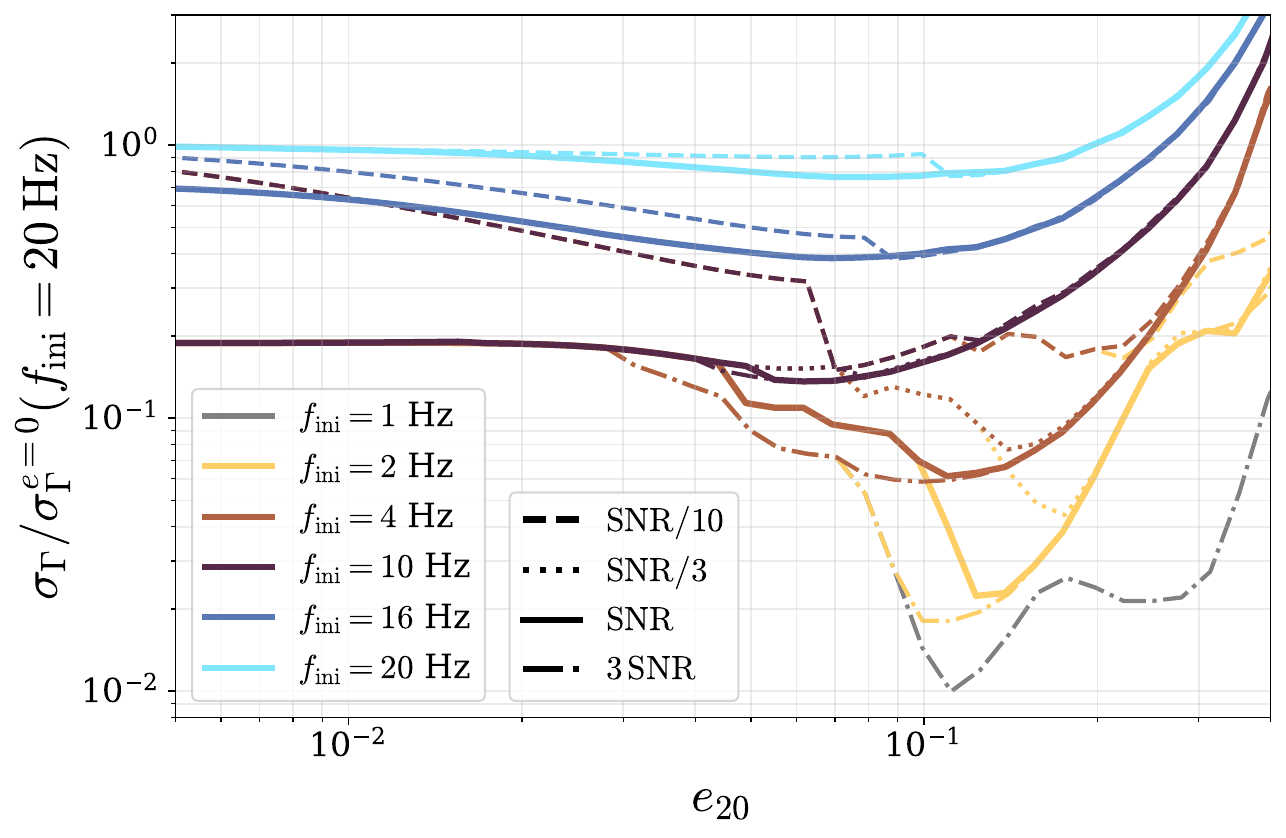}
    \caption{The relative change in the measurement error of $\Gamma$, $\sigma_\Gamma$, with respect to the zero-eccentricity value with $f_{\rm ini}=20$~Hz, as a function of the eccentricity, including the SNR cut, dropping the contribution of harmonics with $(\mathrm{SNR}^n)^2<0.5$. Different values of $f_{\rm ini}$ are shown by different colors, while different line styles mark GW signals with different total SNR.}
    \label{fig:fisher_sigma_SNR}
\end{figure}

Since the effect of the harmonic cut depends on the SNR of the signal, we study its effect on signals with different total SNR.
The results of this analysis are shown in Fig.~\ref{fig:fisher_sigma_SNR}, where the relative change in $\sigma_\Gamma$ with respect to the zero-eccentricity case is shown for signals with different total SNRs, marked by the different line styles. Note that regarding these relative changes, reducing the total SNR of the signal by a given factor is equivalent to increasing the SNR cut threshold by the same factor. As expected, the amplification in the measurability of $\Gamma$ for eccentric binaries is reduced when the total SNR is 3 or 10 times less, however, we can still achieve a maximal improvement of a factor of $\sim20$ or $\sim5$, respectively, by generating the signal with an appropriately low value of $f_{\rm ini}$. On the other hand, with a total SNR 3 times larger than the system we considered in our previous analyses, it is possible to reduce $\sigma_\Gamma$ by a factor of $\sim100$ by increasing the eccentricity to $e_{20}\sim0.1$ and choosing $f_{\rm ini}=1$~Hz. This means that increasing the SNR has two additive effect on the measurability of LSA in eccentric systems: it both reduces $\sigma_\Gamma$ at all eccentricities due to the higher signal power, while it also amplifies the improvement at higher eccentricities by making additional higher harmonics become relevant, thus providing information from even earlier times of the inspiral.

Finally, we note that with a total SNR reduced by a factor of 10, the $(l,m)=(4,4)$ mode is completely discarded from the Fisher analysis  at low eccentricities, creating a similar trend for $f_{\rm ini}=4$~Hz to what is seen in Fig.~\ref{fig:fisher_sigma} for the Fisher results, verifying our previous conclusion.

\subsection{Analysis of GW200105}

We analyze the real event GW200105~\cite{LIGOScientific:2021qlt}\footnote{Strain data were downloaded from the gravitational-wave open science center~\cite{gw200105_data}, while the power-spectral-density (PSD) from the data release in Ref.~\cite{gw200105_psd}.}, for which previous works showed evidence of eccentricity (see, e.g., Ref.~\cite{Morras:2025xfu, Jan:2025fps, Planas:2025plq, Tiwari:2025fua, Kacanja:2025kpr, Phukon:2025cky}), to investigate whether we can infer constraints on LSA. The publicly available cleaned data strain lasts 64~s, therefore we can safely analyze the signal with waveforms generated from 16~Hz or larger. Considering lower values of $f_{\rm ini}$ would require analyzing longer segments of data, which are currently not cleaned. For the analysis we employ the waveform model \texttt{pyEFPEHM} and the multibanding approach for the likelihood evaluation, as for the injection study described above.
\begin{figure}[t]
        \centering
        \includegraphics[width=0.95\columnwidth]{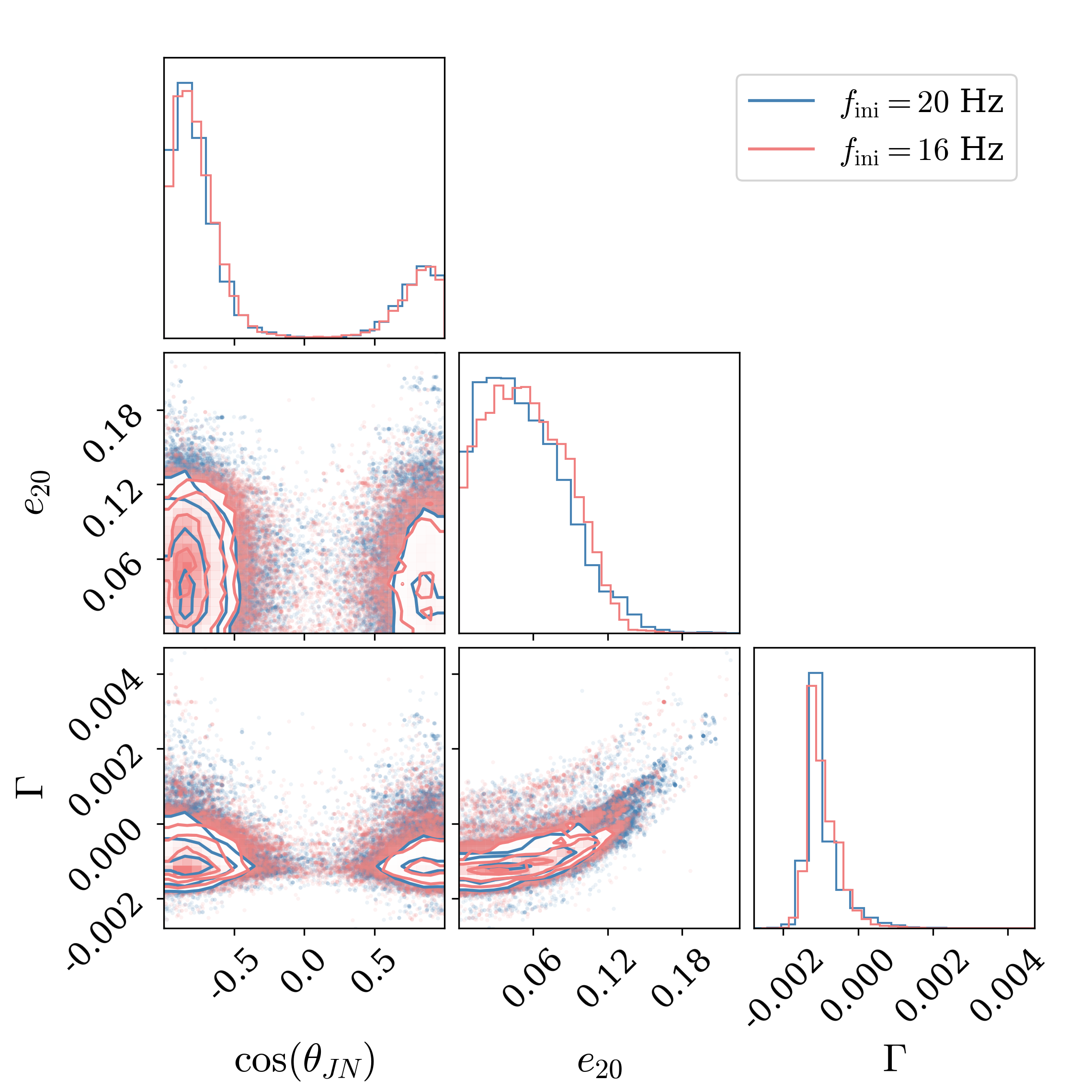}
        \includegraphics[width=0.95\columnwidth]{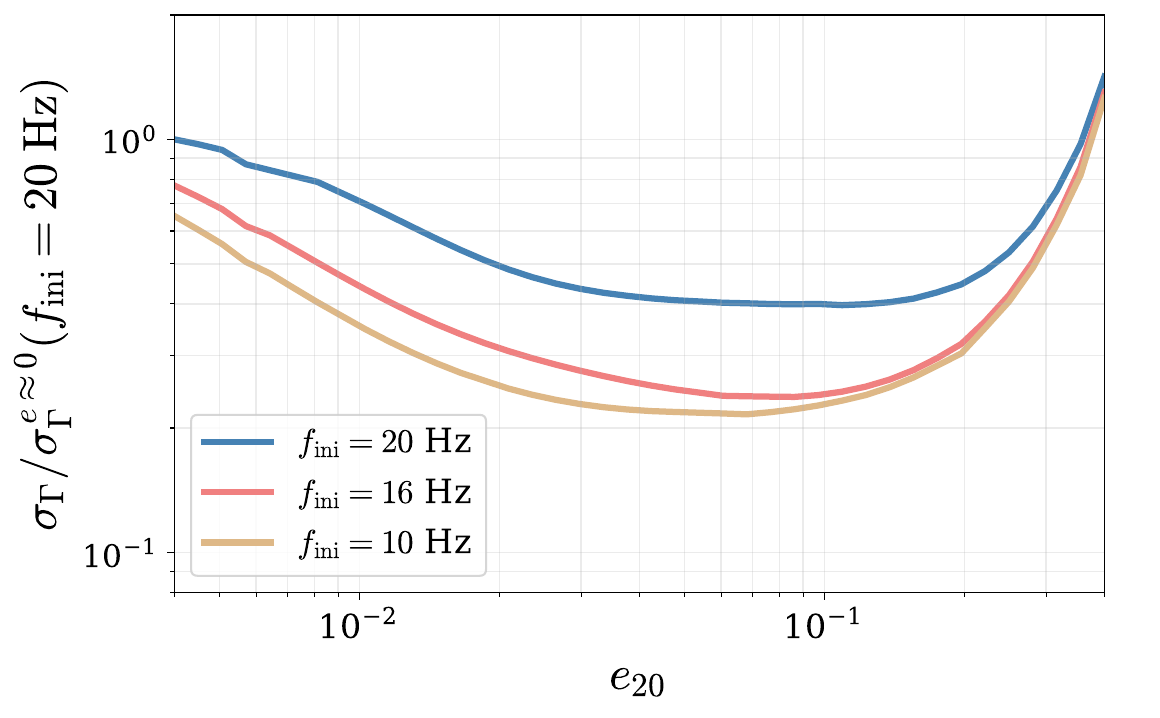}
    \caption{\textit{Top:} Posterior probability distributions for the inclination $\theta_{JN},$ eccentricity $e_{20}$ and the LSA parameter $\Gamma$ recovered from the analysis of GW200105 with the model generated at $f_{\rm ini} = 20\, {\rm Hz}$ and $f_{\rm ini} = 16\,{\rm Hz}$. \textit{Bottom:} Relative change in $\sigma_\Gamma$ with respect to the value at $e_{20}=5\times10^{-3}$ and $f_{\rm ini}=20$~Hz, as a function of the eccentricity, using the Fisher-method for GW200105-like systems, with different values of $f_{\rm ini}$.}
    \label{fig:real_ev}
\end{figure}

The top panel of Fig.~\ref{fig:real_ev} shows the posteriors inferred for the cosine of inclination $\cos (\theta_{JN})$, the eccentricity $e_{20}$ and the LSA parameter $\Gamma$, comparing the case in which the model is generated starting from 20 or 16~Hz. Results are consistent with the ones in Ref.~\cite{Pompili:2026tdf} and Ref.~\cite{Roy:2026mco}, where different waveform approximants were employed in the analysis. Also in our case, the $\Gamma$ posterior peaks at non-zero values: this is due to the strong degeneracy with eccentricity, whose posterior is considerably wider than the one recovered from analyses that do not take into account LSA effects.

The analysis from 16~Hz does not significantly tighten the constraints -- yielding a relative improvement of roughly $10\%$ --, which is expected: changing $f_{\rm ini}$ to 16~Hz we do not even extract the full information from the $n=3$ harmonics. 
The broad eccentricity posterior, peaking at an eccentricity as low as $e_{20}\sim5\times10^{-2}$, hinders the possibility to get a significant increase in the LSA constraint. Looking at Fig.~\ref{fig:fisher_sigma_SNR} as a general reference, the potential improvement at such low eccentricities, which is already quite small for a system with such low SNR as GW200105 -- corresponding roughly to the dashed ${\rm SNR}/10$ lines in the figure --,  would derive mostly from the GW higher-order modes contributions. However, as shown in the posteriors in the top panel of Fig.~\ref{fig:real_ev}, the source emitting GW200105 was most likely a face-on system, for which a large part of higher-order modes contributions is suppressed.

In order to confirm this hypothesis, we calculated the uncertainties for $\Gamma$ for a GW200105-like system using our Fisher-framework. The source masses in this case were set to $m_1=8.9~M_\odot$ and $m_2=1.9~M_\odot$, with $d_L = 280$~Mpc and $\theta_{JN}=0$ set as the luminosity distance and inclination. The bottom panel of Fig.~\ref{fig:real_ev} shows the relative change in $\sigma_\Gamma$ as a function of eccentricity for such systems for different values of $f_{\rm ini}$. This predicts that using $f_{\rm ini}=16$~Hz instead of $20$~Hz shrinks $\sigma_\Gamma$ by $\sim20\%$ at $e_{20}\sim5\times10^{-3}$, while the maximal gain around $e_{20}\sim0.1$ is $\sim40\%$. Considering that the Fisher-method with the previously introduced SNR cut overestimates this improvement for low eccentricities (as shown in Fig.~\ref{fig:fisher_sigma}), we find the results from the full Bayesian analysis of GW200105 to agree with the Fisher predictions reasonably well. 
This Fisher analysis also predicts a further marginal improvement when using $f_{\rm ini}=10$~Hz, and, based on the results shown in Fig.~\ref{fig:fisher_sigma_SNR}, we expect no further improvement beyond this for $f_{\rm ini} = 4 \, {\rm Hz}$; however, analyzing the signal from these lower initial frequencies values would require using $\sim 200$~s or $\sim 2000$~s of data, respectively.

In general, reanalyzing events in search for observable LSA from lower values of $f_{\rm ini}$ would require cleaning longer strains and re-evaluating the PSDs. A more systematic reanalysis of real events with waveform models generated from such lower frequencies is therefore left for future work.

\section{Discussion}
\label{S:Discussion}

Using a state-of-the-art eccentric GW waveform to investigate detectability of LSA in eccentric sources through full Bayesian inference for selected cases, as well as more broadly through Fisher analysis, in this paper we have shown that:
\begin{enumerate}[(i)]
    \item The detectability of the LSA-induced phase shift can be significantly enhanced for moderately eccentric binaries with $e_{20}\sim0.1$, and this enhancement -- more than an order of magnitude for the systems we investigated -- persists in the full PE setup. This effect is due to higher harmonics entering the band at a much earlier time than the $n=2$ harmonic.
    \item In order to fully capture the information from higher eccentric harmonics, one has to initialize the waveform when the first detectable eccentric harmonic enters the detector band, while the frequency of the $n=2$ harmonic might still be below the sensitive band of GW detectors. This equivalently means that a significantly longer time segment of the waveform should be analyzed. We have shown how ignoring this condition can lead to worse constraints on $\Gamma$, similar to the results of previous studies \cite{Roy:2026mco,Pompili:2026tdf,Ansari:2026nhb}.
    \item One can expect an improvement in LSA constraints up to a factor $\sim5$ or more for systems like the ones investigated here with an SNR of $\sim10$, while this increase can be as large as a factor of $\sim50$ for systems with an SNR of $\sim100$.
\end{enumerate}

These results have several important implications. Firstly, note that an increase in SNR doubly strengthens the measurability of LSA in eccentric systems. In addition to simply increasing the precision at all eccentricities, information from further observable harmonics also becomes accessible, amplifying the gain at non-zero eccentricities. Thus an order of magnitude increase in SNR for a system with $e_{20}\sim0.1$ can increase LSA measurement precision by two orders of magnitude instead of just one. 
This suggests very promising perspectives for LSA measurement with third-generation detectors, like Einstein Telescope (ET) \cite{ET:2025xjr}. For such high-SNR signals, one might have to include even further harmonics in eccentric waveform models to fully exploit all the available information from the early evolution of the binary.

Secondly, such a strong amplification in LSA measurability is achieved despite strong parameter correlations at eccentricities $e_{20}\gtrsim0.1$. In case these correlations can be broken, e.g. through the inclusion of higher-order LSA effects in the waveform, a further increase of a factor of few could be achieved.

Thirdly, the measurability of LSA has a very direct correspondence with the length of the analyzed signal containing measurable information, due to the phase shift given by Eq.~\eqref{eq:DPsi_LSA}. This means that if one was to measure LSA of the order $10^{-8}$~s$^{-1}$, corresponding to a typical acceleration in a nuclear cluster, then the analysis of a waveform with $\sim1000$~seconds would become inevitable. This necessitates the development of waveform generation methods even more efficient than the ones currently available for eccentric binaries, as well as fast and accurate likelihood-estimation techniques.

Finally, LSA is only one example of a broad class of astrophysical environmental effects that imprint characteristic dephasing on the GW signals. Since most environmental effects are strongest during the early inspiral, and therefore exhibit steep negative frequency scalings, they should similarly benefit from the harmonic enhancement associated with eccentric binaries. This may substantially improve constraints on the properties of the source environment, including surrounding gas, external tidal fields, or nearby massive companions. More broadly, our results demonstrate that eccentricity and environmental effects should be analyzed jointly. Once eccentricity has been established in a candidate signal, it should be re-analyzed with waveform models including environmental effects, as the harmonic structure can enhance their detectability by orders of magnitude.

\section*{Data availability}

The data from the analyses in this work are available in Zenodo at \href{https://doi.org/10.5281/zenodo.24087524}{\textsf{10.5281/zenodo.24087524}}~\cite{data_release}.

\section*{Acknowledgments}

We thank Lorenzo Pompili for his helpful comments and suggestions as internal LIGO reviewer.
J.T. acknowledges support from the Alexander von Humboldt Foundation under the project no. 1240213 - HFST-P.
A.P. acknowledges funding from the EU Horizon under ERC Starting Grant, no.\ SMArt-101076369. Views and opinions expressed are those of the authors only and do not necessarily reflect those of the European Union or the European Research Council. Neither the European Union nor the granting authority can be held responsible for them. 
G.M.'s research is supported in part by the European Research Council (ERC) Horizon Synergy Grant “Making Sense of the Unexpected in the Gravitational-Wave Sky” grant agreement no. GWSky–101167314.
L.Z. is supported by the European Union’s Horizon 2024 research and innovation program under the Marie Sklodowska-Curie grant agreement No. 101208914.

This research has made use of data or software obtained from the Gravitational Wave Open Science Center (gwosc.org), a service of the LIGO Scientific Collaboration, the Virgo Collaboration, and KAGRA. This material is based upon work supported by NSF's LIGO Laboratory which is a major facility fully funded by the National Science Foundation, as well as the Science and Technology Facilities Council (STFC) of the United Kingdom, the Max-Planck-Society (MPS), and the State of Niedersachsen/Germany for support of the construction of Advanced LIGO and construction and operation of the GEO600 detector. Additional support for Advanced LIGO was provided by the Australian Research Council. Virgo is funded, through the European Gravitational Observatory (EGO), by the French Centre National de Recherche Scientifique (CNRS), the Italian Istituto Nazionale di Fisica Nucleare (INFN) and the Dutch Nikhef, with contributions by institutions from Belgium, Germany, Greece, Hungary, Ireland, Japan, Monaco, Poland, Portugal, Spain. KAGRA is supported by Ministry of Education, Culture, Sports, Science and Technology (MEXT), Japan Society for the Promotion of Science (JSPS) in Japan; National Research Foundation (NRF) and Ministry of Science and ICT (MSIT) in Korea; Academia Sinica (AS) and National Science and Technology Council (NSTC) in Taiwan.

\appendix

\section{Derivation of line-of-sight acceleration phase shift}
\label{app:high_order_LSA}

Perturbations in the waveform due to LSA are purely an observational effect, originating from the delay between the GW being emitted in the source frame and observed in the detector frame. The relation between the source and detector time and frequency, which correspond to the same GW phase $\phi_\mathrm{src}(t_\mathrm{src})=\phi_\mathrm{det}(t_\mathrm{det})$, reads:
\begin{equation}
    f_\mathrm{det} = \frac{f_\mathrm{src}}{1+\Gamma t_\mathrm{src}} \: ,
\end{equation}
\begin{equation}
    t_\mathrm{det} = t_\mathrm{src} + \frac{\Gamma}{2}t_\mathrm{src}^2 \: ,
\end{equation}
where the times are synchronized such that $t_\mathrm{src}=0$ corresponds to $t_\mathrm{det}=0$, and we dropped the term corresponding to an initial non-zero velocity, since it can be absorbed through a shift in the chirp mass. The time domain phase shift due to acceleration becomes:
\begin{align}
    \Delta\phi(t) &= \phi_\mathrm{det}(t) - \phi_\mathrm{src}(t) \nonumber\\ &\approx \phi_\mathrm{det}(t_\mathrm{det}) + \frac{\mathrm{d}\phi_\mathrm{det}}{\mathrm{d}t} \Delta t - \phi_\mathrm{src}(t) \nonumber \\
    &\approx -\pi f \Gamma t^2 \: ,
\end{align}
where $t\equiv t_\mathrm{src}$ and $f \equiv f_\mathrm{src}$. Then, using the first-order accurate relation $\Delta \psi(f) = -\Delta \phi(t(f))$ \cite{Takatsy:2025bfk}, we get
\begin{equation}
    \Delta\psi(f) = \pi \Gamma f t(f)^2 \: .
\end{equation}

Note that the above relation, although leading order in $\Gamma$, is accurate up to arbitrary PN order, provided the appropriate PN-accurate formula is used for $t(f)$. This insight can substantially simplify otherwise lengthy derivations of $\Delta \psi$ for higher PN orders (c.f. equations in Refs.~\cite{Vijaykumar2023-qg,Tiwari:2025aec}).

One can easily see that the above argument generalizes to waveforms with a complex structure of eccentric harmonics and higher-order modes, leading us to the following formula:
\begin{equation}
    \Delta\psi_j(f) = \pi \Gamma f t_j(f)^2 \: ,
\end{equation}
where $j$ represents the indices $( l,m,n )$ of the spherical and eccentric harmonic decompositions.

\begin{table*}[!thb]
\begingroup
\caption{Injection parameters for the additional tests described in Appendix~\ref{sec:further_tests}, compared to the reference system used throughout this work, whose parameters are listed in Sec.~\ref{ssec:setup}. $a_{1,2}$ are the components' dimensionless spin magnitudes, with $\theta_{1,2}$ being the respective zenith angles with respect to the orbital angular momentum; $\phi_{12}$ denotes the difference between the azimuthal angles of the individual spin vectors, and $\phi_{jl}$ the angle between total and orbital angular momentum.
The last column points to the figure with the results for the specific injection.}
\label{tab:test_injections}
\renewcommand*{\arraystretch}{2}
\setlength{\tabcolsep}{6pt}
\begin{tabular}{c c c c c}
\hline
    $m_1$, $m_2$ [$M_\odot$] & $a_1$, $a_2$ & $\theta_1$, $\theta_2$, $\phi_{12}$, $\phi_{JL}$ & $\Gamma$ & Figure \\
        \hline
    [20, 20] & [0.0, 0.0] & [0., 0., 0., 0.] & 0.0 & Reference \\ {}
    [20, 20] & [0.4,0.2] & [1.05, 0.52, 0.1, 0.2] & 0.0 & Fig.~\ref{fig:test_spin} \\ {}
    [41.83, 10.46] & [0.4,0.2] & [1.05, 0.52, 0.1, 0.2] & 0.0 & Fig.~\ref{fig:test_spin_mratio} \\ {}
    [10, 10] & [0.0, 0.0] & [0., 0., 0., 0.] & 0.0 & Fig.~\ref{fig:test_masses} \\ {}
    [20, 20] & [0.0, 0.0] & [0., 0., 0., 0.] & 0.001 & Fig.~\ref{fig:test_alsa} \\

\end{tabular}
\endgroup
\end{table*}

Unlike some previous studies \cite[\eg][]{Roy:2026mco}, here we neglect amplitude corrections. Due to the scope of the present paper, here we do not deliver a detailed derivation but provide a basis for our choice based on the general structure of the correction. After expanding to first order in $\Gamma$, the frequency-domain amplitude of the waveform with LSA takes the following form:
\begin{equation}
    \mathcal{A}_\mathrm{LSA}(f) \approx \mathcal{A}_\mathrm{vac}(f)\left[ 1+ \alpha\Gamma t(f) + \beta\Gamma \frac{f}{\dot{f}}\right] \: ,
\end{equation}
where $\mathcal{A}_\mathrm{vac}$ is the ``vacuum" amplitude without LSA, and $\alpha, \beta$ are scalar factors of $\mathcal{O}(1)$. At leading order $f/\dot{f}=-8/3\, t(f)$. Thus, the amplitude correction is proportional to $\Gamma t(f)$, which is suppressed by a factor of $[t(f)f]^{-1}$ compared to the first-order accurate phase correction, which, at leading order, is:
\begin{equation}
    \left| [t(f)f]^{-1}\right| = \frac{256\pi}{5} (\pi f \mathcal{M})^{5/3} \: ,
\end{equation}
meaning that the amplitude correction might only become comparable to the phase correction near the merger frequency, where the phase shift itself is the smallest. During earlier phases of the inspiral, when contributions from the phase shift can become significant, amplitude corrections are substantially suppressed.

Lastly, let us mention that the above Newtonian approach only remains valid when $\Gamma t \ll 1$, which physically corresponds to the binary velocity remaining in the non-relativistic regime.

\section{Additional tests}
\label{sec:further_tests}

\begin{figure}[tbh]
        \centering        
        \includegraphics[width=0.9\linewidth]{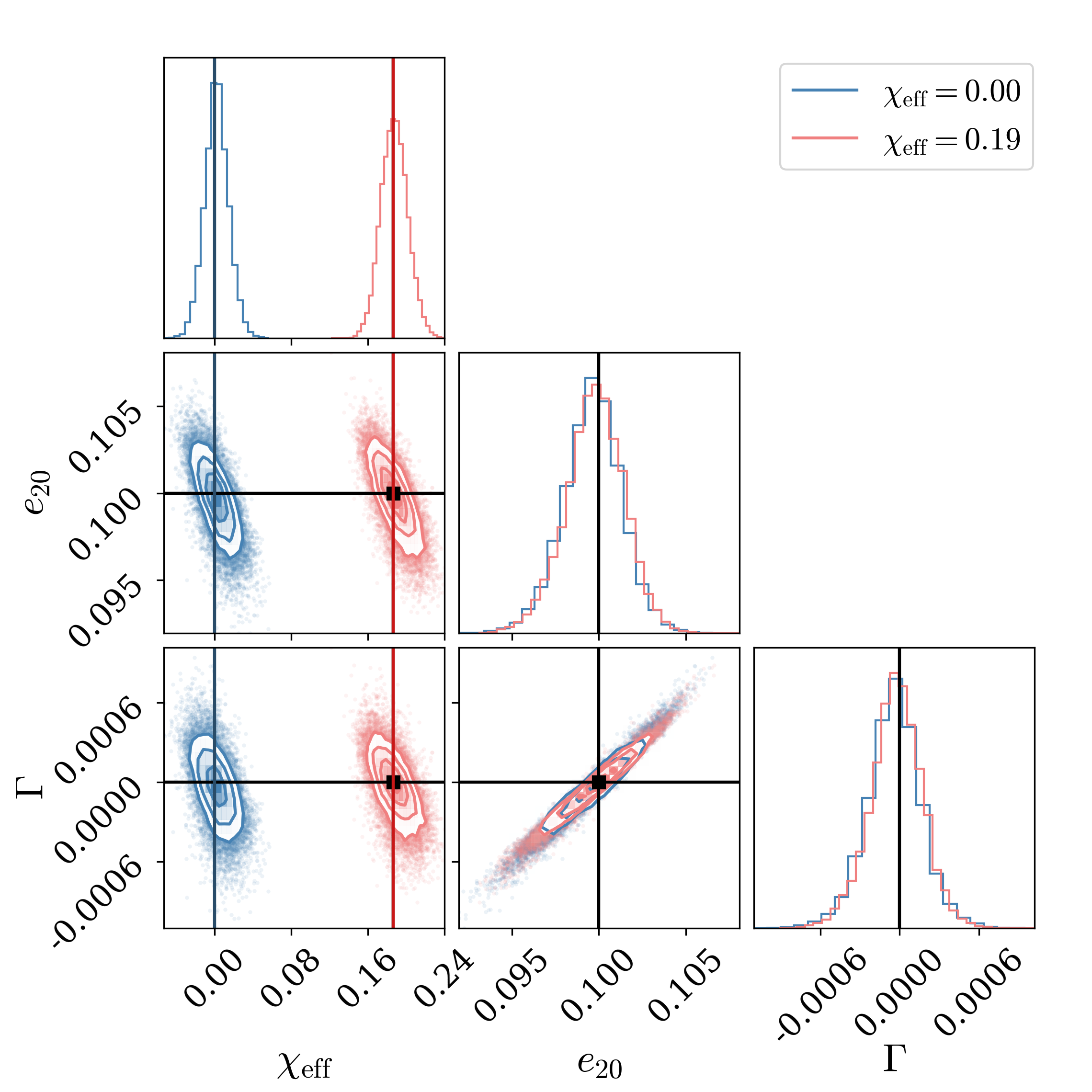}
    \caption{Parameter posteriors recovered for a spinning, precessing system, compared to the reference one. The vertical lines indicate the injected values.}
    \label{fig:test_spin}
\end{figure}

The choice of the parameters of the systems simulated throughout this work was oriented towards simplicity. 
In this section, we show the robustness of the framework against different, often more realistic, choices. All the analyses described below are performed with $f_{\rm ini} = 4 \, {\rm Hz}$, and with an eccentricity at 20~Hz of $e_{20} = 0.1$; the specific injected parameters are listed in Table~\ref{tab:test_injections}.

\begin{figure}[tbh]
        \centering        
        \includegraphics[width=0.9\linewidth]{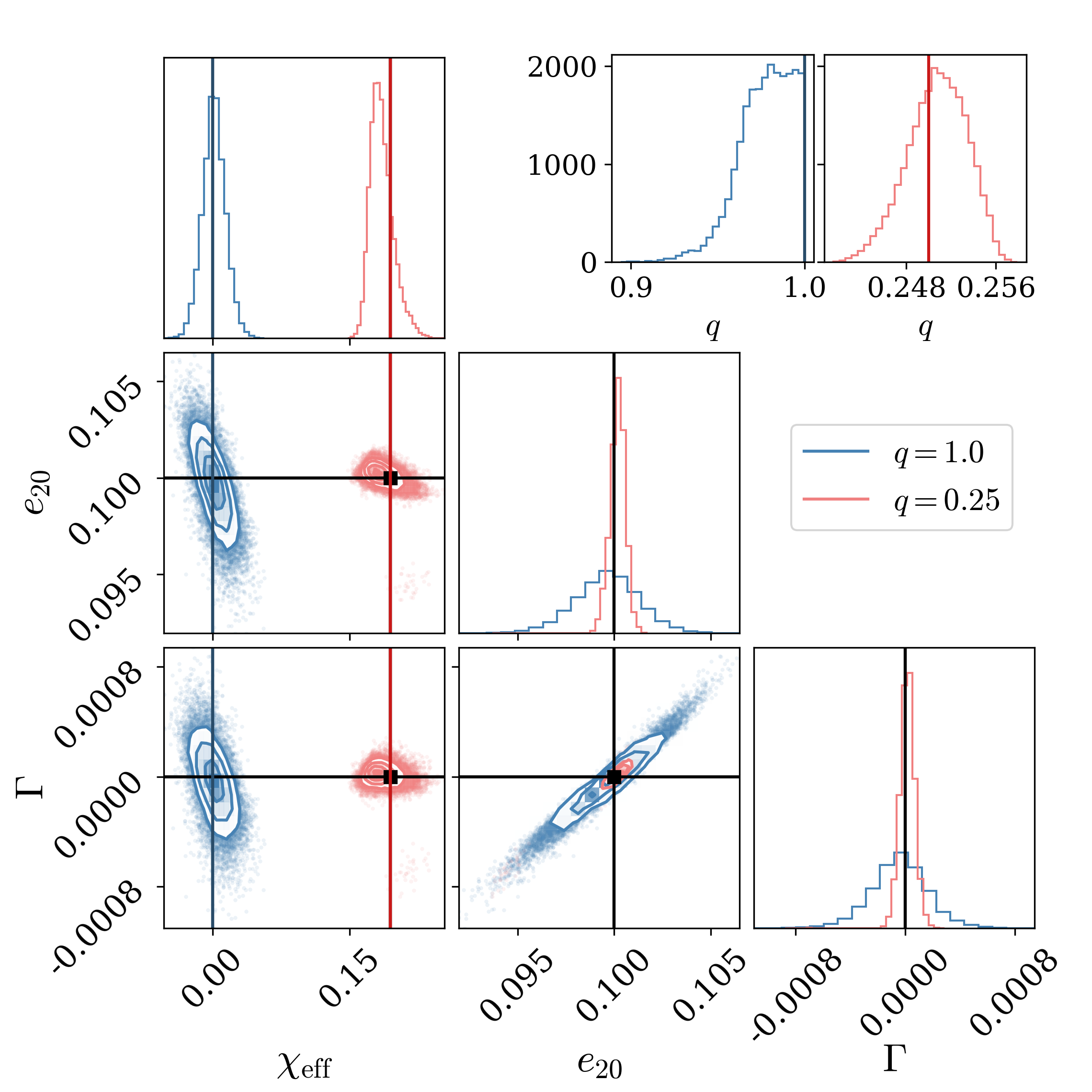}
    \caption{Parameter posteriors obtained for a precessing system with mass ratio $q=0.25$, compared to the reference one. The vertical lines indicate the injected values.}
    \label{fig:test_spin_mratio}
\end{figure}

Fig.~\ref{fig:test_spin} shows the results obtained for a spinning system: all the parameters are well recovered, and spin does not bias the recovery of the other parameters, including eccentricity and $\Gamma$, whose posteriors reproduce the ones obtained in the non-spinning case.

\begin{figure}[tbh]
        \centering        
        \includegraphics[width=0.9\linewidth]{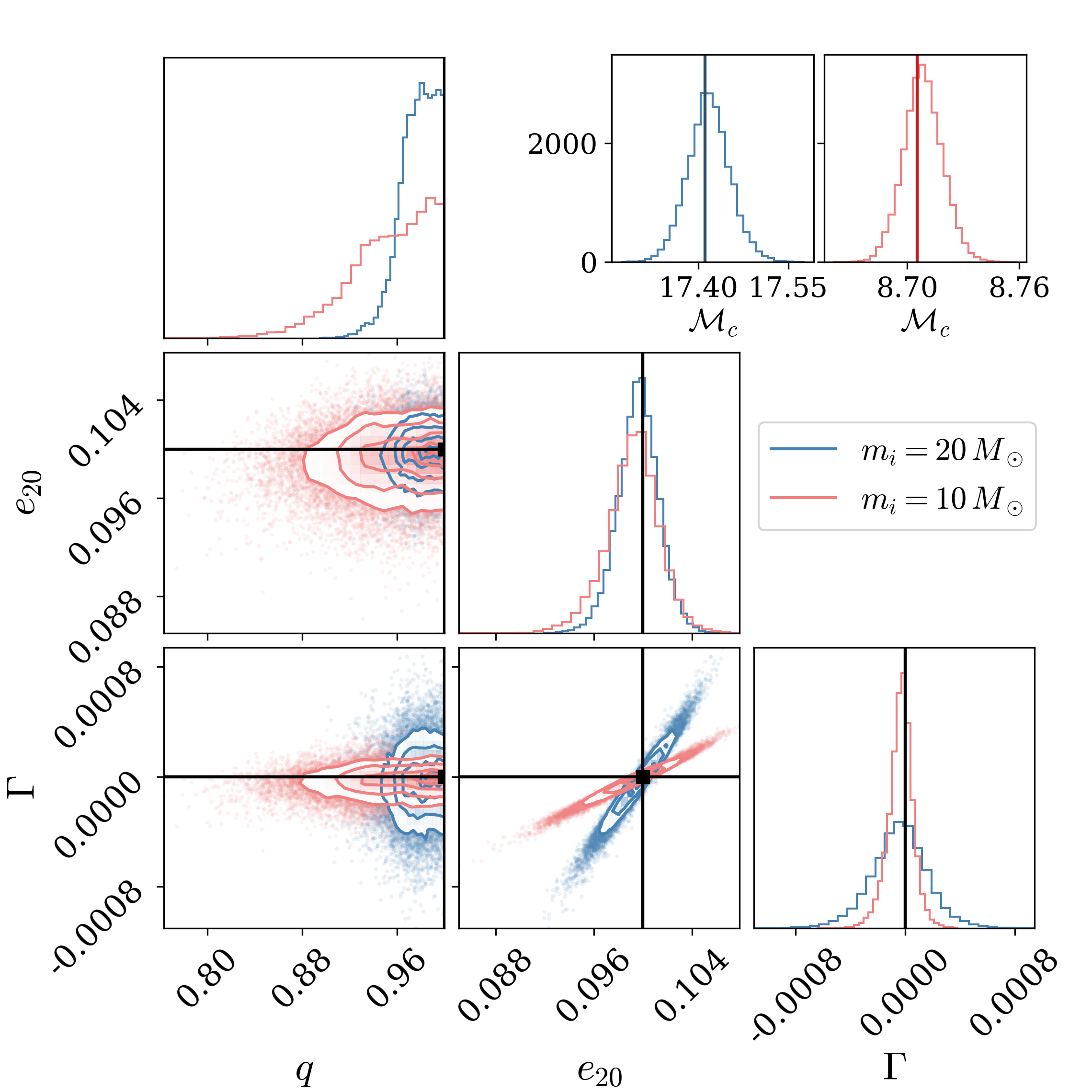}
    \caption{Parameter posteriors recovered for a simulated system with component masses $m_1 =m_2 = 10 \, M_\odot$, compared to the reference one. The vertical lines indicate the injected values.}
    \label{fig:test_masses}
\end{figure}

In Fig.~\ref{fig:test_spin_mratio}, we simulate the signal emitted by a spinning and unequal-mass system, with the same chirp mass as our reference configuration: in this case, the additional information from GW higher-order-modes yields an improvement in the constraints of all the parameters, as shown also in Ref.~\cite{Roy:2026mco}.

\begin{figure}[tbh]
        \centering        
        \includegraphics[width=0.9\linewidth]{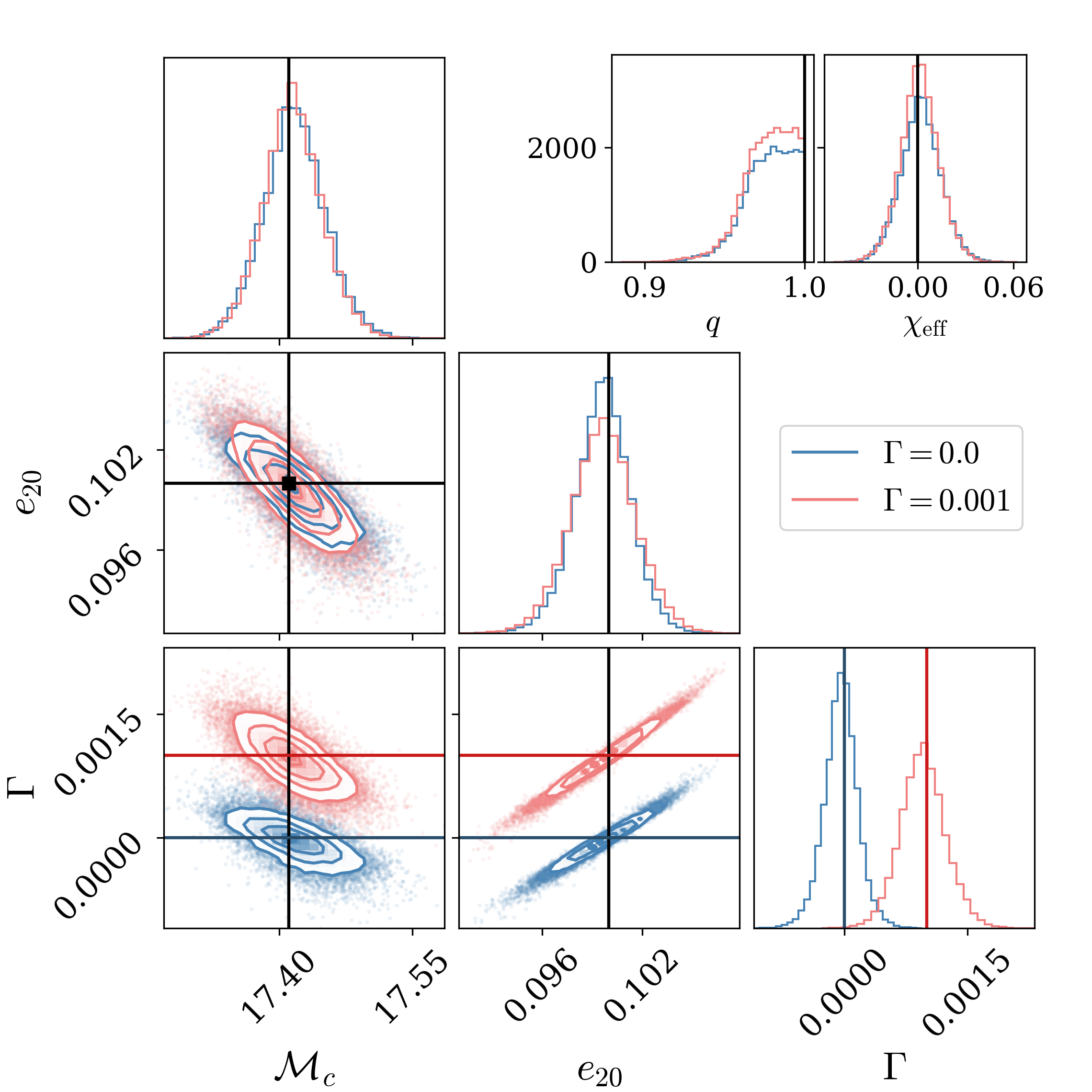}
    \caption{Parameter posteriors recovered for a simulated system with $\Gamma = 0.001$, compared to the reference one. The vertical lines indicate the injected values.}
    \label{fig:test_alsa}
\end{figure}

The mass values for the systems simulated throughout this paper were chosen in order to reduce the computational cost of the analysis. However, we expect systems with lower mass to carry more information since their signals spend more time in band. This is shown in Fig.~\ref{fig:test_masses}, where we compare the results for a system with $m_1 = m_2 = 20 \, M_\odot$, i.e., the reference system in this work, with a system with $m_1 = m_2 = 10 \, M_\odot$, keeping all the other parameters the same.

\begin{figure}[tbh]
        \centering        
        \includegraphics[width=0.9\linewidth]{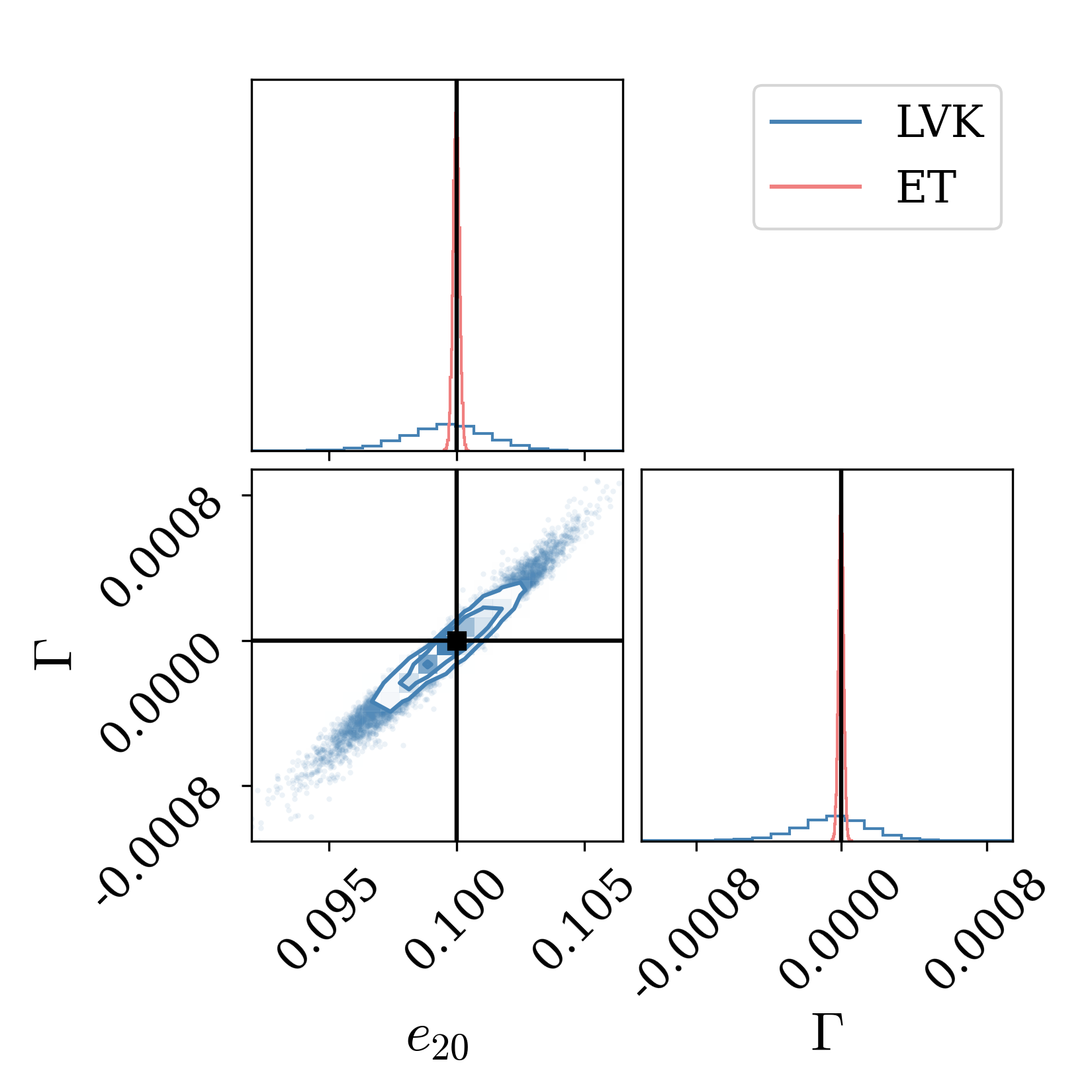}
    \caption{Parameter posteriors recovered for our reference system simulated in an LVK and ET network. The vertical lines indicate the injected values.}
    \label{fig:test_et}
\end{figure}

So far, we always injected signals with $\Gamma = 0$. However, if this environmental effect is present, $\Gamma$ will have a non-zero value. To investigate how the analysis would behave in this case, we simulate a signal with $e=0.1$, and $\Gamma = 0.001$, and the other parameters as in the reference configuration described in Sec.~\ref{ssec:setup}. This specific value of $\Gamma$ lies outside the $90\%$ confidence interval obtained from the reference analysis with $e_{20} = 0.1$ and $f_{\rm ini} = 4$~Hz -- as shown in Figs.~\ref{fig:ecc_diff}-~\ref{fig:fini_diff} --, and would therefore indicate whether we can confidently recover non-zero values of $\Gamma$. Fig.~\ref{fig:test_alsa} shows not only that non-zero values of $\Gamma$ do not bias the recovery of the other parameters of the system, but also that we can correctly recover the injected value of $\Gamma$.

Finally, we investigate the improvements achievable with next-generation detectors, and in particular with ET~\cite{ET:2025xjr, Branchesi:2023mws, et_psd}. Thanks to the technological and design improvements, ET is expected to gain sensitivity well below the one of current detectors, down to a few Hz. This will allow us to access all the information from the early evolution of the systems of interest, leading to massively tighter constraints on their parameters, including $\Gamma$. This is evident from the posterior shown in Fig.~\ref{fig:test_et}, for which we analyzed the detector data from $f_{\rm low} = 5$~Hz, while the starting frequency for waveform generation is kept to $f_{\rm ini}=4$~Hz.

\section{Testing the multibanding likelihood for eccentric signals}
\label{sec:mb_tests}

\begin{figure}[tbh!]
        \centering        
        \includegraphics[width=1.0\linewidth]{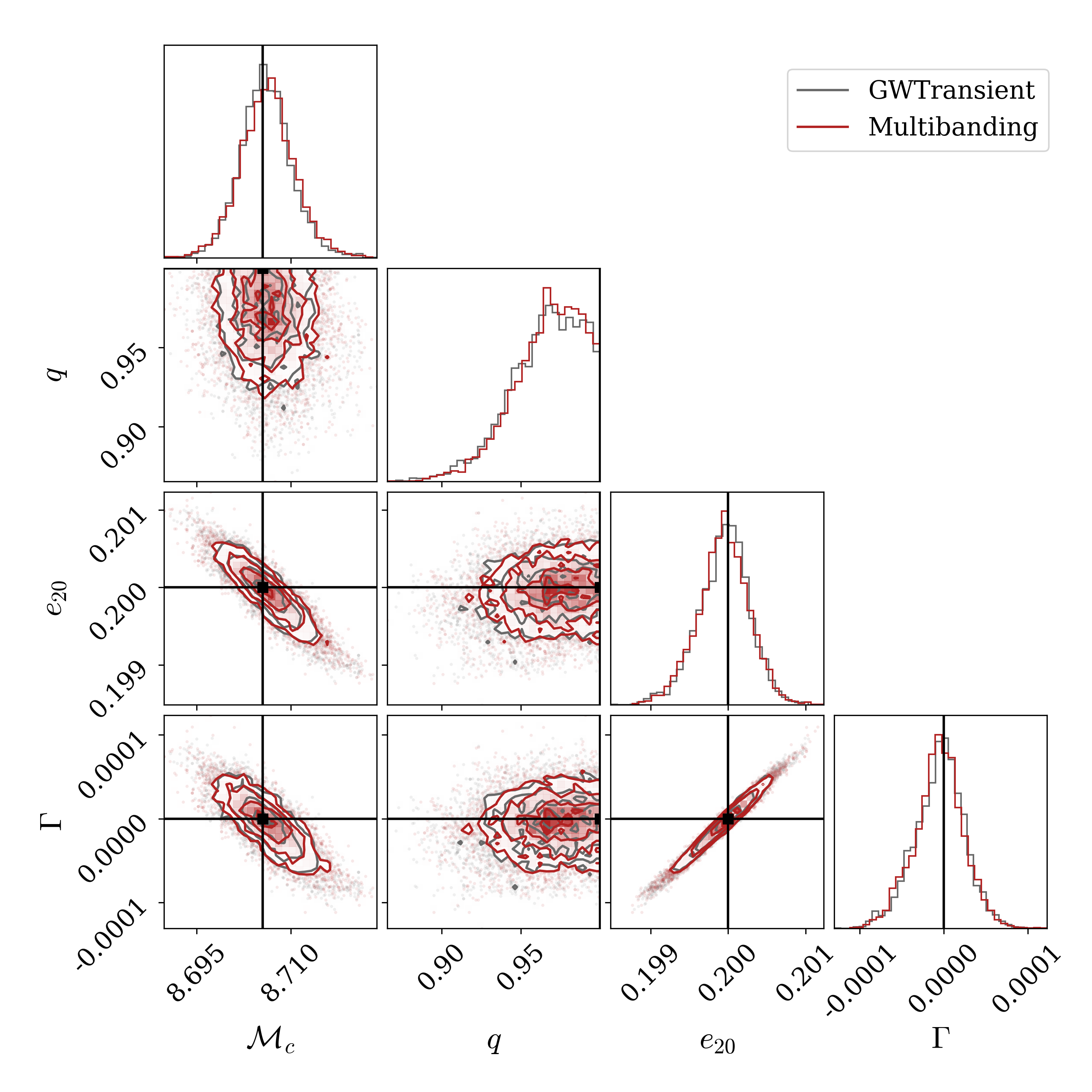}
    \caption{Comparison between the parameter posteriors obtain when analyzing the same simulated signal with the full likelihood, in gray, and with the multibanding approach, in red. Black, vertical lines show the injected values.}
    \label{fig:mb_corner}
\end{figure}

In order to verify that employing the multibanding approach to likelihood computation for eccentric signals does not introduce biases, we perform an analysis on a simulated non-spinning system with $m_1 = m_2 = 10 \, M_\odot$, $e_{20}=0.2$, and $\Gamma=0$, using both the multibanding likelihood and the ``full", non-approximated one (i.e., the \texttt{GWTransient} likelihood in \textsc{Bilby}) . Results are shown in Fig.~\ref{fig:mb_corner}, where we see that the analysis performed with the multibanding likelihood can accurately reproduce the results obtained with the full one.

\begin{figure}[tbh]
        \centering        
        \includegraphics[width=1.0\linewidth]{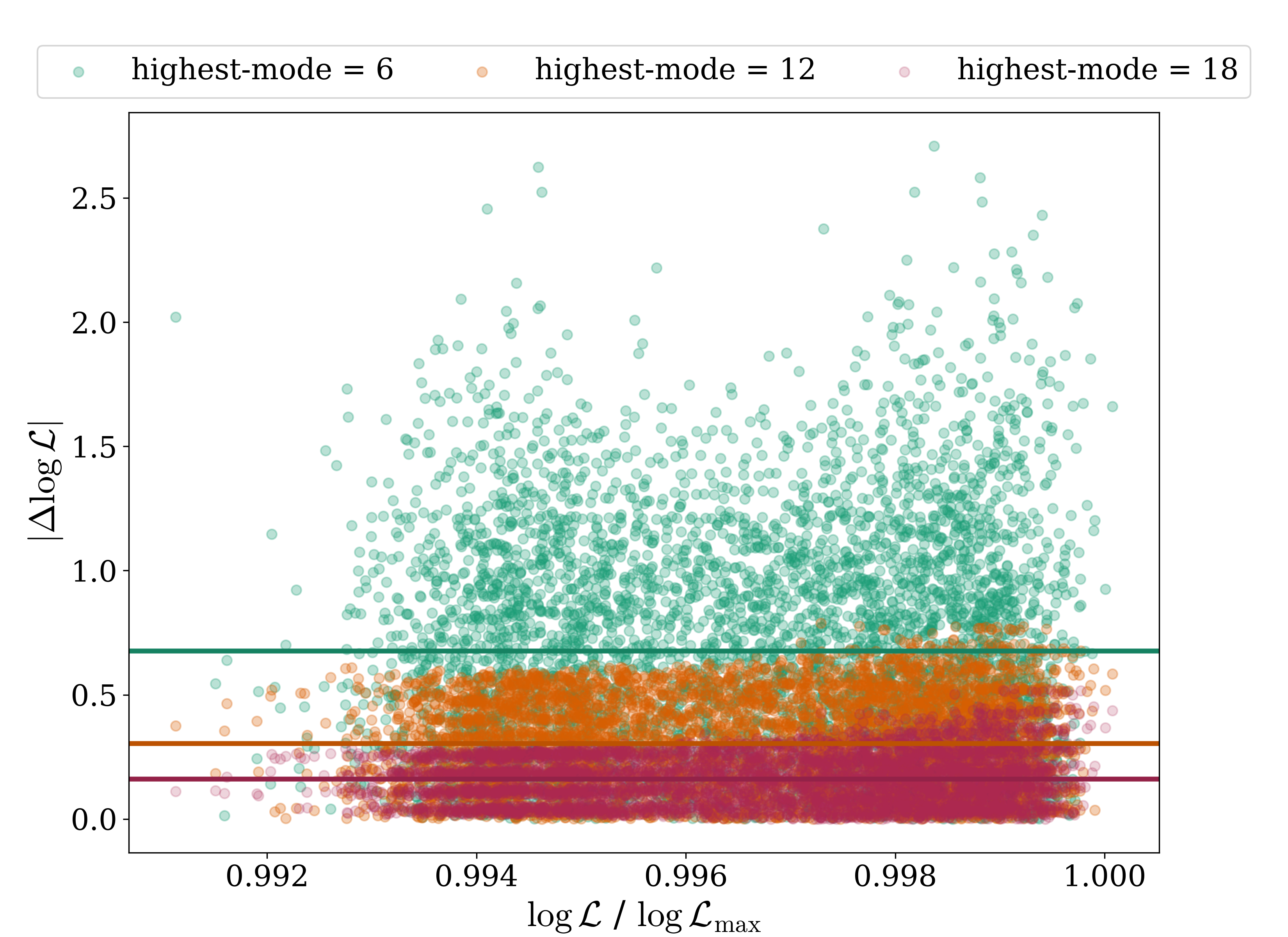}
    \caption{Difference between the full likelihood and the multibanding one, computed with different values for the \textit{highest-mode} setting and with $f_{\rm ini} = 2 \, {\rm Hz}$ for a subset of the samples obtained in our reference run, as function of the full likelihood normalized to the maximum likelihood value in the samples. The horizontal lines indicate the average likelihood difference for each value of \textit{highest-mode}.}
    \label{fig:lik_comp}
\end{figure}

Moreover, we test different settings for the multibanding approach, to ensure that the likelihood estimation is sufficiently close to the non-approximated one. In particular, we focus on the \textit{highest-mode} setting: although it refers to the highest GW higher-order mode considered in the construction of the adaptive frequency grid, it indirectly affects the number of eccentric harmonics that we can capture. This might produce relevant effects especially when $f_{\rm ini} = 2 \, {\rm Hz}$, given the larger number of harmonics included in the signal. Fig.~\ref{fig:lik_comp} shows the difference between the multibanding likelihood and the full one computed for a subset of 5000 samples among the ones collected for the analysis of our reference system when generating the waveform from $f_{\rm ini}=2$~Hz. Overall, we find a very small difference in likelihood also when setting highest-mode to 6, which confirms the goodness of the multibanding approximation. 
However, we see how the error reduces when going to larger values of highest-mode, which, in turn, we expect to slightly influence the constraints on the parameters, including $\Gamma$. Therefore, in the analyses with $f_{\rm ini} = 2 \, {\rm Hz}$, we set highest-mode to 18.

\bibliography{main}

\end{document}